\pdfoutput=1
\documentclass{article}

\usepackage{arxiv}

\usepackage[utf8]{inputenc} %
\usepackage[T1]{fontenc}    %
\usepackage{hyperref}       %
\usepackage{url}            %
\usepackage{booktabs}       %
\usepackage{amsmath,amsfonts,amssymb}
\usepackage{amsthm}
\usepackage{nicefrac}       %
\usepackage{microtype}      %
\usepackage{graphicx}
\usepackage[numbers]{natbib}
\usepackage{doi}

\usepackage{subcaption}
\usepackage{rotating}       %
\usepackage{enumitem}
\usepackage{diagbox}
\usepackage{makecell}

\usepackage{cleveref}       %

\newtheorem{proposition}{Proposition}
\newtheorem{corollary}{Corollary}

\author{
  \href{https://orcid.org/0000-0001-7584-7685}{\includegraphics[scale=0.06]{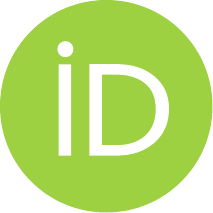}\hspace{1mm}Jing~Liu}%
  \thanks{These authors contributed equally to this work.} \\
  \texttt{st082130@student.spbu.ru}
  \And
  \href{https://orcid.org/0000-0003-3976-7180}{\includegraphics[scale=0.06]{orcid.pdf}\hspace{1mm}Elena~Parilina}%
  \footnotemark[1] \\
  \texttt{e.parilina@spbu.ru}
  \And
  \href{https://orcid.org/0000-0002-1166-7578}{\includegraphics[scale=0.06]{orcid.pdf}\hspace{1mm}Chi~Zhao}%
  \footnotemark[1]\phantom{$^{*}$}\thanks{Corresponding author.} \\
  \texttt{chi.zhao@spbu.ru}
  \AND
  \mdseries Saint Petersburg State University \\
  7/9 Universitetskaya nab., Saint Petersburg, 199034, Russia
}

\hypersetup{
  pdfauthor={Jing Liu, Elena Parilina, Chi Zhao},
}

\title{Leverage-Enhanced Sampling Owen Method for Explainable Artificial Intelligence}
\renewcommand{\shorttitle}{Leverage-Enhanced Sampling Owen Method}
\hypersetup{pdftitle={Leverage-Enhanced Sampling Owen Method for Explainable Artificial Intelligence}, pdfkeywords={Explainable Artificial Intelligence, Shapley Value, Multilinear Extension, Leverage Score Sampling}}

\begin{document}
\maketitle

\begin{abstract}
Recently, the Shapley value has been widely applied in machine learning for feature importance analysis and model explainability. The multilinear extension of games proposed by Owen (1972) expresses the Shapley value as the integral of partial derivatives along the main diagonal of the cube. Direct evaluation of this integral requires exponential-time computation of the partial derivatives. To improve computational efficiency, we propose a sampling-based estimation strategy within Owen's multilinear extension framework, combining stratified sampling with coalition leverage scores. We prove that the resulting estimator is an unbiased estimator of a semivalue, derive its weights and variance in closed form, and show that these weights emphasize marginal contributions to the smallest and the largest coalitions.
	Experimental results demonstrate that the proposed method substantially reduces computational costs while maintaining explanation accuracy.
\end{abstract}


\section{Introduction}\label{sec:intro}

Machine learning models have achieved outstanding predictive performance across a wide range of fields, including medicine \cite{esteva2017skin}, finance \cite{heaton2017deep}, and engineering \cite{shi2025survey}, leading to significant breakthroughs in disease diagnosis, risk prediction, and industrial forecasting \cite{taha2025machine}. As a result, machine learning systems are increasingly being deployed to support critical real-world decisions. However, many high-performance models, such as gradient-boosted trees and deep neural networks, are difficult to understand and explain. Although they typically generate accurate predictions, their decision-making processes are often not transparent. This lack of transparency can reduce users' trust and raise serious concerns in fields where explanations are required. Explainable Artificial Intelligence (XAI) aims to address these issues by making model behaviour more understandable to humans \cite{arrieta2020xai}. Among the most widely used XAI techniques are feature importance methods, which assign an importance score to each input feature based on its influence on the model's prediction, thereby helping users understand why the model makes particular decisions.

Among feature importance methods, game-theoretic methods are particularly popular because they provide a fair and mathematically supported measure of feature importance. Within this framework, a machine learning model is associated with a cooperative game where features represent players, and the characteristic function $v(S)$ measures the model's performance given feature subset $S$. The widely used Shapley value is the unique solution satisfying the axioms of efficiency, symmetry, dummy player, and additivity, and it distributes the total model output fairly among all features by averaging their marginal contributions across all possible coalitions \cite{shapley1953value}. Due to these properties, many Shapley-based methods have been developed for XAI, including SHAP \cite{lundberg2017shap} and ShapG \cite{zhao2025shapg}. Widely used SHAP \cite{lundberg2017shap} provides a unified framework for feature attribution based on the Shapley value. TreeSHAP  runs efficiently on tree-based models but cannot be applied to neural networks or other black-box models \cite{lundberg2020treeshap}. SamplingSHAP \cite{castro2009polynomial,strumbelj2010efficient,lundberg2017shap} is model-agnostic but requires a large number of samples to produce stable explanations. KernelSHAP \cite{lundberg2017shap} is also model-agnostic, but it is computationally slow and has a high variance \cite{covert2021improving}.  LeverageSHAP  is a light-weight modification of KernelSHAP that employs leverage score sampling to provide more accurate Shapley value estimates \cite{musco2024provably}.

However, computing the exact Shapley value is computationally challenging because evaluating the contribution of a single feature requires evaluating all $2^n$ feature subsets over $n$ features, making exact computation infeasible within a reasonable time when the number of features is large. Owen's multilinear extension \cite{owen1972multilinear} offers an alternative formulation by expressing the Shapley value as an integral of partial derivatives along the main diagonal, converting the combinatorial sum into a one-dimensional integral. Note that Owen's multilinear extension is different from the Owen value solution \cite{owen1977}, which is a solution concept for games with predefined group structures rather than a reformulation of the Shapley value as an integral. Although this formulation simplifies the problem in theory, the computational complexity is still exponential. Okhrati and Lipani \cite{okhrati2020multilinear} were the first to turn this formulation into a sampling estimator, called Owen sampling, whose bias and unbiased variants are analysed in \cite{pollmann2026importance}. Unlike these methods, we allocate the sampling budget across the quadrature points proportionally to the Bernoulli variance and, after stratifying coalitions by size at each point, reweight the strata by inverse coalition counts. We should mention that recently developed O-Shap method \cite{zhou2026oshap} computes attributions using the Owen value \cite{owen1977} by constructing feature hierarchies via semantics-aware spatial segmentation, but its design and evaluation focus on vision tasks. Although both O-Shap and our new method use the Owen's results, they use two different solutions \cite{owen1972multilinear,owen1977}.

Our motivation comes from global feature importance, where $v(S)$ is the predictive performance of a model retrained on the feature subset $S$. Every evaluation of $v$ is then a full training run, so the cost of an attribution method is driven mainly by the number of distinct coalitions it evaluates. Such attributions are mostly used to rank features, for example for feature selection or model diagnosis, and the property that matters in this use is deletion faithfulness: removing the features that the method ranks highest should degrade the retrained model the most \cite{samek2017evaluating,hooker2019benchmark}. Deletion faithfulness and closeness to the exact Shapley value are different criteria. In our experiments the exact Shapley ranking is the most faithful one in only one of the four settings where it can be computed (Table~\ref{tab:scores}), and in data valuation, semivalues other than the Shapley value have likewise been found to perform better on downstream tasks \cite{kwon2022beta}. Our method is therefore designed to need few evaluations of the characteristic function while remaining anchored to Owen's representation of the Shapley value, and we evaluate it by deletion faithfulness.

In this paper, we propose a \textit{Sampling Owen method}, a sampling-based algorithm that combines stratified sampling with coalition leverage scores within Owen's multilinear extension framework to compute Shapley-type feature attributions efficiently. Our method is organized as follows. First, we select equally spaced quadrature points on $[0,1]$ and stratify coalitions by size at each point. Sampling based on Bernoulli distribution concentrates almost all samples at coalition sizes near its expected value since the number of coalitions of each size grows combinatorially. Instead, we weight each stratum by the inverse coalition count, a quantity that has the form of the leverage scores in the regression formulation of the Shapley value \cite{musco2024provably}; at each point this moves the sampling effort towards the smallest or the largest coalitions. Within each stratum, coalitions are sampled uniformly and reweighted by importance weights derived from the Bernoulli distribution of coalition sizes. Finally, Simpson's rule is used for approximate integration across all quadrature points. We prove that the resulting estimator is an unbiased estimator of a semivalue \cite{dubey1981value}, a weighted sum of marginal contributions whose weights depend only on coalition size, and we give these weights in closed form. These weights favor the smallest and the largest coalitions, which reduces computational costs while maintaining explanation accuracy.

The main contributions of this work are as follows:
\begin{itemize}
    \item We propose a stratified sampling method within Owen's multilinear extension framework that weights each stratum by the inverse coalition count, a quantity of the same form as leverage scores. This weighting shifts the sampling effort at every quadrature point from the binomial mode towards small or large coalitions; these strata are small, can be enumerated, and are shared by all features, so the number of evaluations of the characteristic function stays low.
    \item We introduce an importance-sampling reweighting procedure for the partial derivative and apply Simpson's rule to integrate it. We prove that the resulting estimator is an unbiased estimator of a semivalue, give its weights and its variance in closed form, and show that it returns the exact Shapley value for additive games, such as a linear model explained at a point.
    \item We conduct experiments on three datasets, including one with 121 features, and show that Sampling Owen method produces explanations with high deletion faithfulness while requiring substantially less computation time than exact methods. Furthermore, it demonstrates competitive performance relative to existing XAI methods and lies on the Pareto frontier of running time and explanation quality in four of the six cases.
\end{itemize}

The structure of this paper is as follows. Section \ref{sec2} reviews feature importance XAI methods and presents alternatives to the Shapley value. Section \ref{sec3} introduces the necessary background on cooperative games, defines the Shapley value and Owen's multilinear extension. Section \ref{sec:method} presents the algorithm of  the proposed Sampling Owen method in detail and derives the properties of the estimator. Section \ref{sec:experiments} shows and analyzes the experimental results. Section \ref{sec:conclusion} briefly concludes.

\section{Feature Importance Methods for XAI}\label{sec2}
\subsection{Existing XAI Methods}\label{subsec:nongame}

Non-game-theoretic methods assign feature importance without relying on cooperative game theory, including tree-based feature importance \cite{breiman2001random,friedman2001gradient}, Permutation Feature Importance (PFI) \cite{fisher2019pfi}, and LIME \cite{ribeiro2016lime}, which fits a local surrogate model around each prediction. These methods are generally fast but often lack axiomatic fairness guarantees.

Most game-theoretic methods assign feature importance based on the Shapley value. SHAP \cite{lundberg2017shap} provides a unified framework and its variants include TreeSHAP \cite{lundberg2020treeshap} for tree models, SamplingSHAP and KernelSHAP \cite{lundberg2017shap} for model-agnostic approximation, and LeverageSHAP \cite{musco2024provably}, which reduces estimator variance via leverage-score-based sampling. Owen sampling \cite{okhrati2020multilinear} estimates the Shapley value from Owen's multilinear extension \cite{owen1972multilinear} by discretizing the integral on a uniform grid and averaging Bernoulli-sampled marginal contributions at each grid point; it is shown to be biased in \cite{pollmann2026importance}, where unbiased variants stratifying either the interval $[0,1]$ or the coalition sizes are proposed. ShapG \cite{zhao2025shapg} constructs an undirected feature graph based on correlation between features and limits the computation range to feature-node neighbourhoods, significantly reducing computational complexity. An improved graph construction and modified XAI method are  proposed in \cite{zhao2025complete}. A detailed description of these methods can be found in \cite{zhao2025shapley}.

\subsection{Alternatives to the Shapley Value in XAI}\label{subsec:efficient}

The Shapley value was introduced by Shapley in \cite{shapley1953value} and it satisfies efficiency, symmetry, additivity, and the dummy player (or null player) properties. But the high computational cost of the Shapley value has led to the development of alternative attribution methods applicable to XAI. As alternatives, the center of imputation set (CIS), the equal-surplus solution (ES), and Fair-Efficient-Symmetric-Perturbation value (FESP) can be considered since they all require only $O(n)$ characteristic function evaluations, and are commonly used as efficient baselines. Existing Additive Feature  Attribution methods (AFA) reformulate the attribution problem as a weighted least-squares problem over all coalitions, resulting in exponential computational complexity of the feature importance calculation problem, similar to that of the exact Shapley value computation. We briefly describe the mentioned values or solutions since we use them for comparison with our proposed XAI method. We remind that all the values are defined for a cooperative  game given by $(N,v)$, where $N$ is the set of players (features) and $v: 2^N\longrightarrow \mathbb{R}$ is a characteristic function assigning weight $v(S)$ for any coalition $S\subset N$ (and $v(S)$ is interpreted as a performance of the model trained on subset of features $S$).

\textit{The CIS (Center of Imputation Set)} proposed in \cite{driessen1991coincidence} is defined as follows:
\begin{equation}
    CIS_i(v) = v(\{i\}) + \frac{1}{n}\left[v(N) - \sum_{j \in N} v(\{j\})\right], \quad \forall\, i \in N.
\end{equation}
The CIS uses only each player's individual payoff and distributes the remaining surplus equally between players, requiring only $n + 1$ characteristic function evaluations, making it very efficient. It satisfies efficiency and symmetry, but it ignores the values of any intermediate coalitions between singletons $\{i\}$, $i\in N$ and the grand coalition $N$, so the contribution of players into intermediate coalitions is not considered while calculating the final solution. In experimental results, we label CIS as Exact CIS.

\textit{The ES (Equal Surplus solution)} \cite{driessen1991coincidence} distributes the surplus equally among players after allocating each player their independent payoff from the grand coalition value:
\begin{equation}
    ES_i(v) = v(\{i\}) + \frac{v(N) - v(\emptyset) - \sum_{j \in N} v(\{j\})}{n}, \quad \forall\, i \in N.
\end{equation}
In cooperative game theory, the characteristic function usually satisfies this condition: $v(\emptyset)=0$ implying that ES coincides with CIS. However, in more general
XAI settings, the value of the empty coalition is often defined by a baseline prediction or quality of the model, e.g., $v(\emptyset)=\mathbb{E}[\hat f(X)]$, which is generally nonzero, so ES and CIS may be different. In our experiments, however, the characteristic function is defined as a predictive-performance score ($R^2$ for regression and $F_1$ for classification) rather than a model output. Accordingly, $v(\emptyset)$ is the performance of a featureless baseline predictor evaluated under the same metric. Hence $v(\emptyset)=0$, and consequently the feature-importance rankings produced by ES and CIS are identical across all datasets and models.

\textit{The FESP (Fair-Efficient-Symmetric-Perturbation) solution} \cite{condevaux2022fair} is based on two extreme feature coalitions, the individual feature value $v(\{j\})$ and the occlusion value, i.e., the value of the grand coalition excluding an individual $v(N\setminus\{j\})$:
\begin{equation}
\varphi_j^{\mathrm{FESP}}(v)
= w\, v(\{j\}) + (1-w)\bigl(-v(N\setminus\{j\})\bigr),
\end{equation}
where the first component $w\,v(\{j\})$ captures the individual marginal feature contribution, while the second component $(1-w)(-v(N\setminus\{j\}))$ represents the contribution of occlusion. The weight $w$ balances these two components and is computed using the characteristic function values:
\begin{equation}
w = \frac{v(N) + \displaystyle\sum_{k \in N} v(N\setminus\{k\})}
{\displaystyle\sum_{k \in N} v(\{k\}) + \sum_{k \in N} v(N\setminus\{k\})}.
\end{equation}
FESP satisfies efficiency, symmetry, and fair treatment, and requires $2n+1$ characteristic function evaluations. 

\textit{The AFA (Additive Feature Attribution) Methods} \cite{hiraki2024afa} provide a unified kernel-based framework in which feature attributions are defined as the solution of a weighted least-squares problem. The choice of kernel $\pi(S)$, the weight assigned to coalition $S$, determines the attribution method. Specifically, the attribution vector $\phi = (\phi_1,\ldots,\phi_n)$ is the solution of the constrained problem
\begin{align}\label{afa_wls}
\phi \in &\operatorname*{arg\,min}_{\psi \in \mathbb{R}^n} \ \sum_{\emptyset \neq S \subsetneq N} \pi(S)\Bigl[\, v(S) - v(\emptyset) - \sum_{i \in S}\psi_i \Bigr]^{2}\\
&\text{subject to} \quad \sum_{i \in N}\psi_i = v(N) - v(\emptyset).\notag
\end{align}
Thus, each coalition $S$ other than $\emptyset$ and $N$ provides one observation: the additive model $\sum_{i\in S}\psi_i$ is fitted to $v(S)-v(\emptyset)$ with weight $\pi(S)$, and the attributions are required to sum to $v(N)-v(\emptyset)$, which is the efficiency condition. The kernel is assumed to satisfy the two conditions imposed in \cite{hiraki2024afa}: (i)  normalization, $\sum_{S \subseteq N \setminus \{i,j\}} \pi(S \cup \{i\}) = 1$ for all $i \neq j$, and (ii)  symmetry, $\pi(S) = \pi(T)$ whenever $|S| = |T|$. Under these conditions the problem has the explicit solution
\begin{equation}\label{afa_gen}
\phi_j = \sum_{\substack{S \ni j \\ S \neq N}} \pi(S)\, v(S) + \frac{v(N) - v(\emptyset) - \displaystyle\sum_{i \in N}\sum_{\substack{S \ni i \\ S \neq N}} \pi(S)\, v(S)}{n}.
\end{equation}
Formula \eqref{afa_gen} expresses the attribution as a function of the kernel, thereby defining a method for any kernel satisfying these two conditions. Formulae \eqref{afa_pren}--\eqref{afa_exp} are obtained by evaluating \eqref{afa_gen} at the three kernels listed below, each of which satisfies both conditions (i) and (ii). The resulting attributions can be computed directly, without numerical optimization, at a computational cost determined by evaluating $v(S)$ for all $2^n$ coalitions.
We use three methods in our experiments:
\begin{itemize}
    \item \textit{LS Prenucleolus} uses a flat kernel $\pi(S) = 1/2^{n-2}$ for $S \neq N$, assigning equal weight to every coalition regardless of size. The resulting attribution coincides with the Least-Squares prenucleolus, a well known  solution in cooperative game theory \cite{ruiz1996least}:
    \begin{equation}
        \label{afa_pren}
        \phi_i^{LS}(v) = \frac{1}{2^{n-2}} \sum_{\substack{S \ni i \\ S \neq N}} v(S) + \frac{v(N) - v(\emptyset) - \displaystyle\frac{1}{2^{n-2}}\sum_{j \in N}\sum_{\substack{S \ni j \\ S \neq N}} v(S)}{n}.
    \end{equation}

    \item \textit{Linear Kernel} uses the kernel $\pi(S) = |S|/(n \cdot 2^{n-3})$ for $S \neq N$, assigning weight proportional to the coalition size:
    \begin{equation}
        \label{afa_lin}
        \phi_i^{LnK}(v) = \sum_{\substack{S \ni i \\ S \neq N}} \frac{|S|}{n \cdot 2^{n-3}}\, v(S) + \frac{v(N) - v(\emptyset) - \displaystyle\sum_{j \in N}\sum_{\substack{S \ni j \\ S \neq N}} \frac{|S|}{n \cdot 2^{n-3}} v(S)}{n}.
    \end{equation}

    \item \textit{Exponential Kernel} uses the kernel $\pi(S) = 2^{|S|-1}/3^{n-2}$ for $S \neq N$, in which the weight doubles with each unit increase in coalition size:
    \begin{equation}
        \label{afa_exp}
        \phi_i^{ExK}(v) = \sum_{\substack{S \ni i \\ S \neq N}} \frac{2^{|S|-1}}{3^{n-2}}\, v(S) + \frac{v(N) - v(\emptyset) - \displaystyle\sum_{j \in N}\sum_{\substack{S \ni j \\ S \neq N}} \frac{2^{|S|-1}}{3^{n-2}} v(S)}{n}.
    \end{equation}
\end{itemize}

\textit{Semivalues} \cite{dubey1981value} drop the efficiency axiom of the Shapley value and keep linearity, symmetry, positivity and the dummy axiom. A semivalue assigns to player $i$ a weighted sum of its marginal contributions, with weights that depend only on the coalition size:
\begin{equation}\label{semivalue}
\xi_i(v) = \sum_{S \subseteq N\setminus\{i\}} \lambda_{|S|}\bigl[v(S\cup\{i\}) - v(S)\bigr], \qquad \lambda_s \geq 0, \quad \sum_{s=0}^{n-1}\binom{n-1}{s}\lambda_s = 1.
\end{equation}
The Shapley value is the semivalue with $\lambda_s = s!\,(n-s-1)!/n!$, and the Banzhaf value is the one with $\lambda_s = 2^{-(n-1)}$. Both arise from Owen's multilinear extension (Section~\ref{sec3}): the Shapley value integrates the partial derivative of the extension along the diagonal with respect to the uniform distribution on $[0,1]$, and the Banzhaf value evaluates it at $t=1/2$ \cite{owen1975banzhaf}. Replacing the uniform distribution by other probability distributions on $[0,1]$ yields other semivalues \cite{dubey1981value}, such as the Beta Shapley values of Kwon and Zou \cite{kwon2022beta} for Beta distributions. The method of Section~\ref{sec:method} also starts from Owen's representation and estimates a semivalue that departs from the Shapley weights: it gives more weight to both the smallest and the largest coalitions (Section~\ref{subsec:properties}).

In the following section, we introduce a new feature importance method based on Owen's multilinear extension and a leverage-enhanced sampling procedure. In the experimental results we compare our method with four groups of XAI methods described in this section: (1) Shapley value-based methods including TreeSHAP, SamplingSHAP, KernelSHAP, LeverageSHAP, ShapG, and Improved ShapG; (2) Exact Shapley and Owen values if the number of features in the dataset is not too large; (3) XAI methods based on solutions other than the Shapley value, including Exact CIS, ES, and FESP;  (4) AFA methods using  the LS prenucleolus, linear and exponential kernels.

\section{From the Shapley Value to Multilinear Extensions}\label{sec3}

\subsection{The Shapley Value}\label{subsec:shapley}

The Shapley value, introduced by Lloyd Shapley for cooperative games, provides a principled way to ``fairly'' distribute the total payoff among players \cite{shapley1953value}. A cooperative game is a pair $(N, v)$ where $N = \{1, \ldots, n\}$ is the set of players and $v: 2^N \rightarrow \mathbb{R}$ is the characteristic function assigning a value to every coalition $S \subseteq N$, with $v(\emptyset) = 0$. In the XAI context, players correspond to features and $v(S)$ measures the predictive performance of the model using feature subset $S$.

The Shapley value $(\varphi_i(v): i\in N)$ is the unique solution satisfying efficiency, symmetry, dummy, and additivity axioms:
\begin{equation}
    \label{shapley}
    \varphi_i(v) = \sum_{S \subseteq N \setminus \{i\}} \frac{|S|!\,(n - |S| - 1)!}{n!}
    \bigl[v(S \cup \{i\}) - v(S)\bigr].
\end{equation}
Its computation requires $2^n$ evaluations corresponding to any coalition $S \subseteq N$, which is computationally infeasible for large $n$. In the experimental results, the exact computation of the Shapley value is labeled Exact Shapley and it is presented only for datasets with small number of features.

\subsection{Multilinear Extension of a Cooperative  Game}\label{subsec:mle}

The multilinear extension of a cooperative game $(N,v)$, introduced by Guillermo Owen \cite{owen1972multilinear}, lifts the characteristic function from the discrete hypercube $\{0,1\}^n$ to the continuous unit cube $[0,1]^n$. Given $v: 2^N \rightarrow \mathbb{R}$, its multilinear extension $f: [0,1]^n \rightarrow \mathbb{R}$ is defined as follows:
\begin{equation}
\label{eq:mle}
f(x) = \sum_{S\subseteq N} \left[\prod_{j\in S} x_j \cdot \prod_{j\notin S} (1-x_j)\right] \cdot v(S),
\end{equation}
where $x_j \in [0,1]$ represents the probability that player $j$ participates in a coalition.

By Owen's theorem \cite{owen1972multilinear}, the player $i$'s Shapley value component equals the integral of the partial derivative of $f$ along the main diagonal:
\begin{equation}\label{owen val}
\varphi_i = \int_0^1 \frac{\partial f}{\partial x_i}(t, t, \ldots, t) \, dt.
\end{equation}
The partial derivative along the diagonal admits the closed form:
\begin{equation}\label{deriv Owen}
\frac{\partial f}{\partial x_i}(t, \ldots, t) = \sum_{S\subseteq N\setminus\{i\}} t^{|S|} (1-t)^{n-1-|S|} \bigl[v(S\cup\{i\}) - v(S)\bigr],
\end{equation}
which is the expected marginal contribution of player $i$ when each other player joins independently with probability $t$. We write $f_i(t)$ for this partial derivative, so that $\varphi_i=\int_0^1 f_i(t)\,dt$. In the experimental results, the exact Shapley value computed via Owen's multilinear extension is labeled Exact Owen.\footnote{In our implementation of Exact Owen, the partial derivatives are calculated by formula \eqref{deriv Owen}, while the integral in \eqref{owen val} is evaluated numerically on a dense grid, the resulting quadrature error is negligible.}

\section{Leverage-Enhanced Sampling Owen Method (LESOM)}\label{sec:method}

This section describes the proposed Sampling Owen  method and corresponding algorithm in detail. The method consists of seven steps: selecting quadrature points, allocating the coalition samples, stratified sampling by coalition size, computing leverage scores, tilting by Bernoulli probability, importance-weighted estimation of the integrand, and final Simpson integration. There are three default parameters, they are the number of quadrature points $K$, total coalition sample size $m$ and the enumeration threshold $\theta$. We set them as follows:
\begin{enumerate}
\item Let $K = 11$ for $n \leq 20$ features  and $K = 21$ otherwise;
\item We set $m = \min(5n,\, 2^{n-1})$.\footnote{The choice of $m$ is based on empirical experiments for better performance of our algorithm and this value of $m$ is applied for all datasets under consideration in Section 5. The cap at $2^{n-1}$ prevents oversampling when $n$ is small.}
\item We enumerate a stratum if it contains at most $\theta = 100$ coalitions.
\end{enumerate}

\subsection{Algorithm}\label{subsec:algorithm}
\begin{enumerate}
    \item \textbf{Selecting quadrature points.} We use Simpson's rule to numerically approximate the integral $\int_0^1 f_i(t)\,dt$. We select $K$ equally distributed points:
    \begin{equation}
        t_k = \frac{k-1}{K-1}, \quad k = 1, 2, \ldots, K, \qquad h = \frac{1}{K-1},
    \end{equation}
    where $K$ must be odd. The Simpson approximation is:
    \begin{equation}
    \int_0^1 f_i(t)\,dt \approx \frac{h}{3}\sum_{k=1}^{K} c_k f_i(t_k),
    \end{equation}
where $(c_1,c_2,c_3,c_4,\ldots,c_{K-1},c_K)=(1, 4, 2, 4, \ldots, 4, 1)$.

    \item \textbf{Setting coalition count and allocation.} The total coalition sample size is $m = \min(5n,\, 2^{n-1})$. At the two endpoints $t_1 = 0$ and $t_K = 1$, the partial derivative is computed exactly, without sampling (specifically, $f_i(0) = v(\{i\}) - v(\emptyset)$ and $f_i(1) = v(N) - v(N \setminus \{i\})$), so  no coalitions are assigned to these points. All $m$ coalitions  are distributed across the $K-2$ interior quadrature points in proportion to the per-feature Bernoulli variance\footnote{The stochastic variable of appearance of a feature at point $t_k$ obeys Bernoulli distribution with probability of success $t_k$ and probability of failure $1-t_k$, and this variable has an expectation $t_k$ and variance $t_k(1-t_k)$. This distribution is denoted by $\mathrm{Bernoulli}(t_k)$.} $t_k(1-t_k)$, with each interior point guaranteed at least one coalition:
    \begin{equation}
        \label{alloc_points}
        m_k = \max\!\left(1,\; \left\lfloor m \cdot \frac{t_k(1-t_k)}{\sum_{k'=2}^{K-1} t_{k'}(1-t_{k'})} \right\rfloor\right), \quad k = 2, \ldots, K-1.
    \end{equation}
    Any leftover from rounding is distributed to the highest-variance points. The weight $t_k(1-t_k)$ is a simple proxy for the variability of a sampled marginal contribution at $t_k$: the size of a $\mathrm{Bernoulli}(t_k)$ coalition has variance $(n-1)t_k(1-t_k)$, which vanishes at the endpoints, where $f_i$ is computed exactly, and is largest at $t_k=1/2$. Since it depends only on $t_k$, the allocation is fixed before the characteristic function is evaluated.

    \item \textbf{Stratified sampling.}
    At each quadrature point $t_k$, we select $m_k$ coalitions to find an approximated value of $f_i(t_k)$. Coalitions are grouped into strata by size $s$ ($s = 0, 1, \ldots, n-1$). The per-stratum sampling probability $\tilde{p}_s(t_k)$ (defined in Step~5) determines how $m_k$ is split across strata. Define the sampling budget for stratum $s$ at quadrature point $t_k$ as follows:
    \begin{equation}
    \label{alloc_strata}
    m_{s,k} = \lfloor\, \tilde{p}_s(t_k) \cdot m_k \,\rfloor, \, s = 0, 1, \ldots, n-1,
    \end{equation}
    with any remainder distributed among the strata with the largest $\tilde{p}_s$. Within each stratum, if $m_{s,k} \geq \binom{n-1}{s}$, all coalitions of size $s$ are considered, without sampling. Otherwise, $m_{s,k}$ coalitions are sampled uniformly at random (with replacement). All coalitions of size $s$ are also considered, without sampling, when $m_{s,k}\geq 1$ and $\binom{n-1}{s} \leq \theta$. All evaluations are cached across points and players, so the actual evaluation count may differ from $m$. We write $n_{s,k}$ for the number of coalitions of size $s$ used at $t_k$:
    \begin{equation}
    \label{nsk}
    n_{s,k}=
    \begin{cases}
    0, & \text{if } m_{s,k}=0,\\[2pt]
    \binom{n-1}{s}, & \text{if } m_{s,k}\ge 1 \text{ and } \bigl(m_{s,k}\ge\binom{n-1}{s} \text{ or } \binom{n-1}{s}\le\theta\bigr),\\[2pt]
    m_{s,k}, & \text{otherwise.}
    \end{cases}
    \end{equation}
    Like the allocation of Step~2, these numbers depend only on $n$, $K$, $m$ and $\theta$. We call stratum $s$ \textit{enumerated} at $t_k$ in the second case of \eqref{nsk} and \textit{sampled} in the third.

    \item \textbf{Leverage score computation.} Naive $\mathrm{Bernoulli}(t_k)$ sampling draws sizes from binomial distribution $\mathrm{Binomial}(n-1,t_k)$, whose expected value is $(n-1)t_k$ and it visits extreme sizes with small probability. To rebalance effort across sizes, we weight each size by its inverse coalition count,
    \begin{equation}
    \ell_s = \frac{1}{\binom{n-1}{s}}, \quad s = 0, 1, \ldots, n-1,
    \end{equation}
    a quantity of the same form as the leverage scores in the regression view of the Shapley estimation \cite{musco2024provably}. In this regression view, which underlies KernelSHAP, every coalition $S$ with $0<|S|<n$ contributes one row of a constrained weighted least-squares problem, and Musco and Witter \cite{musco2024provably} showed that the statistical leverage score of that row equals $1/\binom{n}{|S|}$. Sampling rows in proportion to their leverage scores therefore gives every coalition size the same total probability. Our $\ell_s$ has the same form for the coalitions $S\subseteq N\setminus\{i\}$ that enter the partial derivative \eqref{deriv Owen}, of which there are $\binom{n-1}{s}$ of size $s$. In our method, $\ell_s$ is used in this sense, as a sampling weight across coalition sizes.

    \item \textbf{Bernoulli tilting.} Since $\ell_s$ is independent of $t_k$, we tilt it by the true size mass $\binom{n-1}{s}t_k^s(1-t_k)^{n-1-s}$ to restore the dependence on $t_k$:
    \begin{equation}
    p_s(t_k) = \ell_s \cdot \binom{n-1}{s} t_k^s (1-t_k)^{n-1-s} = t_k^s (1-t_k)^{n-1-s},
    \end{equation}
    the per-coalition probability, which spreads smoothly across sizes. Since multiplying by $\ell_s$ removes the binomial coefficient, $\{p_s(t_k)\}_{s=0}^{n-1}$ no longer sum to one, so we normalize them into the stratum-selection distribution:
    \begin{equation}
        \label{tildep}
        \tilde{p}_s(t_k) = \frac{t_k^s(1-t_k)^{n-1-s}}{\displaystyle\sum_{s'=0}^{n-1} t_k^{s'}(1-t_k)^{n-1-s'}},
    \end{equation}
    and this probability is used in Steps~3 and~6.

    \item \textbf{Importance-Weighted Estimation.}
    Each coalition of size $s$ selected at $t_k$ receives the importance weight
    \begin{equation}\label{eq:weight}
    w_s(t_k) = \frac{\Pr_{\mathrm{Bernoulli}(t_k)}(|S| = s)}{\tilde{p}_s(t_k)} = \frac{\binom{n-1}{s}t_k^s(1-t_k)^{n-1-s}}{\tilde{p}_s(t_k)},
    \end{equation}
    the ratio of the probability of size $s$ under $\mathrm{Bernoulli}(t_k)$ to the probability $\tilde{p}_s(t_k)$ with which its stratum is selected. Finally, an approximated value of $f_i(t_k)$ is
    \begin{equation}\label{fhat}
    \hat{f}_i(t_k) = \frac{\displaystyle\sum_{j} w_{|S_j|}(t_k) \cdot \bigl[v(S_j \cup \{i\}) - v(S_j)\bigr]}{\displaystyle\sum_{j} w_{|S_j|}(t_k)}.
    \end{equation}
Here $S_j$ runs over the coalitions selected at $t_k$ in Step~3, listed with multiplicity. Section~\ref{subsec:properties} gives the expectation of $\hat{f}_i(t_k)$ and of the final estimate exactly. We should notice that since we are interested in only the feature ranking (not in the absolute values of feature importance), which is invariant to a common additive shift across players, no efficiency normalization is imposed.

    \item \textbf{Simpson Integration.}
    Finally, Simpson's rule integrates the estimates across all quadrature points:
    \begin{equation}
    \hat{\varphi}_i = \frac{h}{3}\sum_{k=1}^{K} c_k\,\hat{f}_i(t_k),\quad i\in N,
    \end{equation}
    where the Simpson coefficients $c_k$, $k=1,\ldots,K$, are defined in Step 1.
\end{enumerate}

\noindent\textbf{Remark.} For $t_k \neq 1/2$ the distribution \eqref{tildep} favors small ($t_k < 1/2$) or large ($t_k > 1/2$) sizes, while at $t_k = 1/2$ it is uniform over all sizes.

In the experimental results, we label this sampling method based on Owen's multilinear extension framework as Sampling Owen.

\subsection[Properties of the Estimator]{Properties of the Estimator}\label{subsec:properties}

In this section we show that the estimator of Section~\ref{subsec:algorithm} is an unbiased estimator of a semivalue~\eqref{semivalue} and calculate its variance in closed form. Write $D_i(S)=v(S\cup\{i\})-v(S)$ for the marginal contribution of player $i$ to a coalition $S\subseteq N\setminus\{i\}$. If coalition $S$ is drawn uniformly at random from all coalitions of size $s$, the marginal contribution of player $i$ is random. Let $\mu_{i,s}$ and $\sigma^2_{i,s}$ be the mean and the variance of $D_i(S)$ over the $\binom{n-1}{s}$ coalitions of size $s$:
\begin{equation}\label{musigma}
\mu_{i,s} = \binom{n-1}{s}^{-1}\sum_{\substack{S\subseteq N\setminus\{i\}\\ |S|=s}}D_i(S), \qquad
\sigma^2_{i,s} = \binom{n-1}{s}^{-1}\sum_{\substack{S\subseteq N\setminus\{i\}\\ |S|=s}}\bigl(D_i(S)-\mu_{i,s}\bigr)^2.
\end{equation}
Grouping the terms of \eqref{deriv Owen} by size gives $f_i(t)=\sum_{s=0}^{n-1}\binom{n-1}{s}t^s(1-t)^{n-1-s}\mu_{i,s}$. Each of these binomial weights integrates to $1/n$ over $[0,1]$, so Owen's formula \eqref{owen val} becomes
\begin{equation}\label{shapley_sizes}
\varphi_i=\frac{1}{n}\sum_{s=0}^{n-1}\mu_{i,s},
\end{equation}
that is, the Shapley value gives every coalition size the same weight $1/n$. The estimator of Section~\ref{subsec:algorithm} also combines coalitions only through their sizes, but with other weights. Let $a_k=(h/3)c_k$ be the Simpson weights, which sum to one, and let
\begin{equation}\label{As}
q_{s,k}=\frac{n_{s,k}\,w_s(t_k)}{\sum_{s'=0}^{n-1}n_{s',k}\,w_{s'}(t_k)}, \qquad A_s=\sum_{k=1}^{K}a_k\,q_{s,k},
\end{equation}
Here $q_{s,k}$ is the share of stratum $s$ in the estimate \eqref{fhat} at the interior point $t_k$, and $A_s$ is the total weight of size $s$ in the final estimate. At the endpoints, where $f_i$ is computed exactly, we set $q_{0,1}=q_{n-1,K}=1$ and all other $q_{s,1}$ and $q_{s,K}$ to zero.

\begin{proposition}\label{prop:estimator}
For fixed $n$, $K$, $m$ and $\theta$, the weights $A_s$ are deterministic, nonnegative and sum to one. For every player $i$:
\begin{enumerate}[label=(\roman*)]
\item $\mathbb{E}[\hat\varphi_i]=\xi_i:=\sum_{s=0}^{n-1}A_s\,\mu_{i,s}$, i.e.\ $\hat\varphi_i$ is an unbiased estimator of the semivalue \eqref{semivalue} with $\lambda_s=A_s/\binom{n-1}{s}$;
\item $\operatorname{Var}(\hat\varphi_i)=\sum_{k=2}^{K-1}a_k^2\sum_{s\in R_k}q_{s,k}^2\,\sigma^2_{i,s}/n_{s,k}$, where $R_k$ is the set of strata that are sampled rather than enumerated at $t_k$.
\end{enumerate}
\end{proposition}

\begin{proof}
Fix a player $i$ and an interior point $t_k$. All coalitions of stratum $s$ enter \eqref{fhat} with the same weight $w_s(t_k)$, so $\hat f_i(t_k)=\sum_s q_{s,k}\bar D_{i,s,k}$, where $\bar D_{i,s,k}$ is the average of $D_i(S)$ over the $n_{s,k}$ coalitions of size $s$ used at $t_k$. If the stratum is enumerated, $\bar D_{i,s,k}=\mu_{i,s}$. If it is sampled, $\bar D_{i,s,k}$ is the mean of $n_{s,k}$ independent uniform draws from the stratum, so its expectation is $\mu_{i,s}$ and its variance is $\sigma^2_{i,s}/n_{s,k}$. At the endpoints, $\hat f_i(t_1)=\mu_{i,0}$ and $\hat f_i(t_K)=\mu_{i,n-1}$.

Summing over the quadrature points gives $\hat\varphi_i=\sum_k a_k\hat f_i(t_k)=\sum_s\sum_k a_kq_{s,k}\bar D_{i,s,k}$. Taking expectations gives (i), and since draws in different strata and at different points are independent, we obtain (ii). The weights $q_{s,k}$ depend only on $n$, $K$, $m$ and $\theta$ (Step~3), are nonnegative and sum to one over $s$, so the $A_s$ are deterministic and nonnegative, and $\sum_s A_s=\sum_k a_k=1$. Finally, $\xi_i=\sum_{S\subseteq N\setminus\{i\}}\bigl(A_{|S|}/\binom{n-1}{|S|}\bigr)D_i(S)$ has the form \eqref{semivalue} with $\lambda_s=A_s/\binom{n-1}{s}$, and $\sum_s\binom{n-1}{s}\lambda_s=\sum_sA_s=1$.
\end{proof}

With the default parameters, most of the weight $A_s$ goes to the smallest and the largest coalition sizes. Their strata are small compared with the middle ones, and a stratum with at most $\theta$ coalitions is enumerated in Step~3, so its contribution to the estimate is exact. The coalitions of these sizes, such as $\{j\}$, $\{i,j\}$, $N\setminus\{j\}$ and $N$, are also the same for every feature and every quadrature point, so each of them is evaluated only once. As a result, the characteristic function is evaluated at far fewer than $2^n-1$ coalitions.

A simple case is a linear model $f(x)=\beta_0+\beta^{\top}x$ explained at a point $x$, with the absent features set to their means $\bar x$. Then $v(S)=f(x_S,\bar x_{N\setminus S})-f(\bar x)=\sum_{j\in S}\beta_j(x_j-\bar x_j)$, so adding feature $j$ to any coalition changes $v$ by the same amount $c_j=\beta_j(x_j-\bar x_j)$. Such games are called additive: $v(S)=\sum_{j\in S}c_j$ for all $S\subseteq N$. The same holds for any additive model $f(x)=\beta_0+\sum_j g_j(x_j)$. In an additive game the weights $A_s$ play no role, as the following corollary shows.

\begin{corollary}\label{cor:additive}
Let $v$ be an additive game. Then $\varphi_i=c_i$, and Sampling Owen returns $\hat\varphi_i=c_i$ for every player $i$ and every realization of the sampling, so the sampling introduces no error. KernelSHAP, LeverageSHAP and the AFA methods of Section~\ref{sec2} also return $c_i$ whenever their weighted least-squares problem has a unique solution, which is always the case for the AFA methods.
\end{corollary}

\begin{proof}
For an additive game, $D_i(S)=c_i$ for every $S\subseteq N\setminus\{i\}$. Hence $\mu_{i,s}=c_i$ and $\sigma^2_{i,s}=0$ for every $s$, and \eqref{shapley_sizes} gives $\varphi_i=c_i$. In Sampling Owen, every marginal contribution in \eqref{fhat} equals $c_i$, so $\hat f_i(t_k)=c_i$ at every quadrature point, including the endpoints. Since $\sum_k a_k=1$, we obtain $\hat\varphi_i=c_i$, whichever coalitions are selected.

KernelSHAP, LeverageSHAP and the AFA methods all minimize $\sum_{S\in\mathcal{C}}\omega_S\bigl(v(S)-\sum_{j\in S}\phi_j\bigr)^2$ subject to $\sum_j\phi_j=v(N)$, with weights $\omega_S>0$. They differ in the collection $\mathcal{C}$ and in the weights. KernelSHAP and LeverageSHAP sample $\mathcal{C}$ at random, while the AFA methods use all coalitions other than $\emptyset$ and $N$, with $\omega_S=\pi(S)$ as in \eqref{afa_wls}. In an additive game, $\phi=(c_1,\ldots,c_n)$ satisfies the constraint and makes every residual zero, so it minimizes the objective. It is the only minimizer if the indicator vectors of the coalitions in $\mathcal{C}$, together with the all-ones vector, span $\mathbb{R}^n$. This always holds for the AFA methods, because $\mathcal{C}$ then contains all singletons.
\end{proof}

We checked this numerically for multiple linear regression models on UCS and Adult Income: at ten test points and with 20 random seeds, the estimates of Sampling Owen, KernelSHAP (without $\ell_1$ regularization) and LeverageSHAP differ from $\beta_j(x_j-\bar x_j)$ only by rounding errors, with relative errors of at most $1.1\times10^{-14}$.

The games in our experiments are not additive, since the models (LightGBM and TabM) are nonlinear and the characteristic function is the $R^2$ or $F_1$ score of a model retrained on each coalition. The methods therefore give different attributions, and we compare them by deletion tests in Section~\ref{sec:experiments}.

\section{Experiments and Results}\label{sec:experiments}

\subsection{Datasets}\label{subsec:datasets}

We evaluate XAI methods on three publicly available datasets covering both regression and binary classification tasks:
\begin{description}
	\item[UCS] (Unconfined Compressive Strength) is a regression dataset containing 437 observations and 15 features. The features include mine tailings, superplasticizer, fine aggregate, coarse aggregates, water, and curing age, etc. The prediction target is the unconfined compressive strength.
	\item[Adult Income] (Adult Census Income) is a binary classification dataset with 30718 observations and 11 features including age, education level, occupation, marital status, relationship, race, sex, capital gain and loss, and hours worked per week, etc. The task is to predict whether an individual's annual income exceeds \$50K.
	\item[Crime] (Communities and Crime) is a regression dataset with 2215 observations and 121 features covering demographic, socioeconomic, housing, and law-enforcement characteristics of US communities, such as racial composition, poverty level, family structure, and housing vacancy, etc. The prediction target is the rate of violent crimes per population. With 121 features, exact Shapley, exact Owen, and AFA methods are computationally infeasible, so only approximation and sampling-based methods are evaluated on this dataset. 
\end{description}

For regression tasks, we use $R^2$ as the characteristic function, and for classification tasks, we use the F1 score. Therefore, for regression (classification) tasks the value of characteristic function $v$ for subset of features $S$ is equal to the determination coefficient $R^2$ (F1 metric) for the model constructed using only features from $S$. 

\subsection{AI models}\label{subsec:AImodels}

For each dataset from Section \ref{subsec:datasets}, we construct two AI predictive models and evaluate feature importance for them:
\begin{itemize}
  \item \textit{LightGBM} \cite{ke2017lightgbm}: A gradient boosting decision tree model that uses histogram-based split finding for efficient training and reduced memory usage.
  \item \textit{TabM} \cite{gorishniy2024tabm}: A deep learning model for tabular data that achieves high performance through parameter-efficient ensembling, enabling multiple ensemble members to be represented within a single neural network while sharing most parameters.
\end{itemize}

\subsection{Evaluation Metrics}\label{subsec:metrics}

To compare the proposed LESOM (Leverage-Enhanced Sampling Owen Method) with other XAI methods described in Section \ref{sec2}, we use three evaluation metrics to compare these XAI methods:
\begin{itemize}
    \item \textit{Feature perturbation:} Each method ranks features by importance. Having ranks given by all XAI methods, we start removing features one by one starting from the most important, retrain the model each step, and check how much performance drops with removing features one by one ($R^2$ for regression and accuracy for classification datasets). A bigger drop in performance means the method identified a more important feature. This procedure is represented on a graph. It is a most-relevant-first deletion test \cite{samek2017evaluating} with retraining, as in the remove-and-retrain benchmark of \cite{hooker2019benchmark}; retraining ensures that the drop reflects the information carried by the removed features rather than the response of a fixed model to inputs it was not trained on. We refer to the quality that it measures as deletion faithfulness.

    \item \textit{Weighted slope $\mathcal{S}$}: This metric is related to the previous item. Based on feature perturbation, let $b_i$ denote the model performance after removing the $i$ most important features. We define a weighted score:
    \begin{equation}\label{slope}
        \mathcal{S} = \sum_{i=1}^{L} \Delta_i  \alpha^{i-1},
    \end{equation}
    where $\Delta_i = b_{i-1} - b_i$ is the performance drop when the $i$-th most important feature is removed, $L$ is the number of removal steps ($L=n-1$ for UCS and Adult Income, and $L=20$ for Crime), $\alpha \in (0, 1]$ is a decay factor giving greater weight to top-ranked features. A larger $\mathcal{S}$ means the ranking better captures most important features. In our work, we set $\alpha = 0.8$.

    \item \textit{Running time:} The running time of each XAI method for generating explanations (includes the time of retraining the model).
\end{itemize}

\subsection{Results and Analysis}\label{subsec:results}

The results are divided in three parts following evaluation metrics described in Section~\ref{subsec:metrics}:
\begin{enumerate}
\item Tables \ref{tab:ucs_lgbm} and \ref{tab:ucs_tabm} present feature importance rankings generated by all considered XAI methods for the UCS dataset under LightGBM and TabM models, respectively. For Adult Income and Crime datasets feature importance rankings are provided in the Appendix.

\item Figures \ref{fig:ucs}--\ref{fig:crime} show the feature removal curves for all three datasets under both AI models (LightGBM and TabM). Each curve shows how model performance changes as features are removed in descending order of importance. A steeper decline in the curve indicates that the XAI method provides more faithful explanations. The weighted slope scores $\mathcal{S}$ (shown in each figure) are summarized in Table \ref{tab:scores}.

\item Running time for all datasets and both AI models is  given in Table \ref{tab:runtime}.
\end{enumerate}

\begin{sidewaystable}[htbp]
\centering
\caption{Feature importance ranking for the UCS dataset (LightGBM). Entries are feature labels.$^{a}$ Merged column headers indicate that the corresponding methods produce identical explanation results}
\label{tab:ucs_lgbm}
\setlength{\tabcolsep}{3pt}
\renewcommand{\arraystretch}{1.1}
\scriptsize
\begin{tabular}{l|c|c|c|c|c|c|c|c|c|c|c}
\toprule
\diagbox[width=5em,height=3em]{Rank}{XAI}
 & \makecell{Sampling\\SHAP\\Kernel\\SHAP}
 & \makecell{Leverage\\SHAP}
 & \makecell{Original\\ShapG}
 & \makecell{Improved\\ShapG}
 & \makecell{Exact CIS\\ES}
 & FESP
 & \makecell{LS\\Prenucleolus}
 & \makecell{Linear\\Kernel}
 & \makecell{Exponential\\Kernel}
 & \makecell{Exact Shapley\\Exact Owen}
 & \makecell{Sampling\\Owen}
\\
\midrule
Top 1  & 1  & 1  & 1  & 1  & 5  & 5  & 1  & 1  & 1  & 1  & 1  \\
Top 2  & 2  & 5  & 5  & 5  & 2  & 1  & 5  & 5  & 4  & 5  & 5  \\
Top 3  & 3  & 2  & 9  & 4  & 1  & 2  & 2  & 2  & 5  & 2  & 2  \\
Top 4  & 4  & 11 & 11 & 9  & 9  & 9  & 9  & 9  & 2  & 9  & 9  \\
Top 5  & 5  & 9  & 7  & 12 & 15 & 15 & 7  & 7  & 3  & 11 & 11 \\
Top 6  & 6  & 15 & 4  & 2  & 11 & 11 & 11 & 11 & 15 & 7  & 7  \\
Top 7  & 7  & 3  & 15 & 13 & 4  & 4  & 15 & 15 & 7  & 15 & 15 \\
Top 8  & 8  & 7  & 2  & 11 & 3  & 3  & 3  & 3  & 6  & 3  & 4  \\
Top 9  & 9  & 12 & 13 & 14 & 7  & 7  & 13 & 4  & 9  & 4  & 3  \\
Top 10 & 10 & 4  & 3  & 15 & 12 & 12 & 12 & 6  & 8  & 12 & 13 \\
Top 11 & 11 & 10 & 14 & 7  & 13 & 13 & 4  & 13 & 11 & 13 & 12 \\
Top 12 & 12 & 13 & 12 & 6  & 14 & 14 & 6  & 12 & 14 & 14 & 14 \\
Top 13 & 13 & 6  & 10 & 3  & 6  & 6  & 14 & 14 & 12 & 6  & 6  \\
Top 14 & 14 & 8  & 6  & 10 & 10 & 10 & 8  & 8  & 13 & 10 & 10 \\
Top 15 & 15 & 14 & 8  & 8  & 8  & 8  & 10 & 10 & 10 & 8  & 8  \\
\bottomrule
\end{tabular}
\par\medskip
{\scriptsize $^{a}$ Feature labels: 1: age; 2: sp; 3: mt\_ssa; 4: mt; 5: water; 6: scm2; 7: opc; 8: scm1; 9: ca; 10: scm1\_ssa; 11: fa; 12: scm2\_ssa; 13: scm2\_pi; 14: scm1\_pi; 15: mt\_pi.}
\end{sidewaystable}

\begin{sidewaystable}[htbp]
\centering
\caption{Feature importance ranking for the UCS dataset (TabM). Entries are feature labels.$^{a}$ Merged column headers indicate that the corresponding methods produce identical explanation results}
\label{tab:ucs_tabm}
\setlength{\tabcolsep}{3pt}
\renewcommand{\arraystretch}{1.1}
\scriptsize
\begin{tabular}{l|c|c|c|c|c|c|c|c|c|c|c}
\toprule
\diagbox[width=5em,height=3em]{Rank}{XAI}
 & \makecell{Sampling\\SHAP}
 & \makecell{Kernel\\SHAP}
 & \makecell{Leverage\\SHAP}
 & \makecell{Original\\ShapG}
 & \makecell{Improved\\ShapG}
 & \makecell{Exact CIS\\ES\\FESP}
 & \makecell{LS\\Prenucleolus}
 & \makecell{Linear\\Kernel}
 & \makecell{Exponential\\Kernel}
 & \makecell{Exact Shapley\\Exact Owen}
 & \makecell{Sampling\\Owen}
\\
\midrule
Top 1  & 1  & 1  & 1  & 1  & 1  & 2  & 1  & 1  & 1  & 1  & 1  \\
Top 2  & 2  & 2  & 2  & 2  & 2  & 1  & 2  & 2  & 2  & 2  & 2  \\
Top 3  & 3  & 3  & 15 & 15 & 15 & 15 & 7  & 7  & 7  & 15 & 15 \\
Top 4  & 4  & 4  & 3  & 11 & 5  & 5  & 3  & 3  & 3  & 3  & 5  \\
Top 5  & 7  & 7  & 7  & 5  & 10 & 11 & 11 & 11 & 4  & 7  & 3  \\
Top 6  & 5  & 5  & 5  & 7  & 11 & 3  & 15 & 15 & 11 & 5  & 11 \\
Top 7  & 15 & 15 & 11 & 3  & 9  & 9  & 5  & 4  & 15 & 11 & 7  \\
Top 8  & 11 & 11 & 4  & 10 & 6  & 10 & 4  & 5  & 5  & 4  & 9  \\
Top 9  & 8  & 8  & 10 & 9  & 3  & 4  & 6  & 6  & 6  & 10 & 10 \\
Top 10 & 14 & 14 & 9  & 6  & 7  & 6  & 10 & 10 & 10 & 9  & 4  \\
Top 11 & 13 & 13 & 6  & 4  & 12 & 7  & 9  & 9  & 9  & 6  & 6  \\
Top 12 & 6  & 6  & 14 & 12 & 4  & 14 & 14 & 14 & 14 & 14 & 14 \\
Top 13 & 10 & 12 & 13 & 14 & 14 & 12 & 8  & 8  & 8  & 12 & 12 \\
Top 14 & 12 & 10 & 8  & 8  & 13 & 8  & 13 & 12 & 13 & 8  & 13 \\
Top 15 & 9  & 9  & 12 & 13 & 8  & 13 & 12 & 13 & 12 & 13 & 8  \\
\bottomrule
\end{tabular}
\par\medskip
{\scriptsize $^{a}$ Feature labels: 1: age; 2: sp; 3: mt\_ssa; 4: mt; 5: water; 6: scm2; 7: opc; 8: scm1; 9: ca; 10: scm1\_ssa; 11: fa; 12: scm2\_ssa; 13: scm2\_pi; 14: scm1\_pi; 15: mt\_pi.}
\end{sidewaystable}

Now we discuss these results in detail.

\textbf{Feature importance ranking.}
We make a detailed analysis of feature ranking for the UCS dataset, while results of feature ranking for Adult Income and Crime datasets can be analyzed in a similar way using Tables \ref{tab:adult_lgbm}--\ref{tab:crime_tabm} provided in the Appendix.

Tables \ref{tab:ucs_lgbm} and \ref{tab:ucs_tabm} present the feature importance rankings generated by different explanation methods on the UCS dataset under LightGBM and TabM, respectively. Overall, most methods consistently identify features 1 (\textit{age}), 2 (\textit{sp}), 11 (\textit{fa}), and 15 (\textit{mt\_pi}) as highly important features.

For the LightGBM model (Table \ref{tab:ucs_lgbm}), Exact Shapley, Exact Owen, and Sampling Owen exhibit highly similar rankings. Sampling Owen matches Exact Owen  for the top-7 ranks (features 1, 5, 2, 9, 11, 7, and 15) and differs only at ranks 8--9, where features 3 and 4 are interchanged, and at ranks 10--11, where features 13 and 12 are interchanged. CIS and ES rank water (5) first, while FESP also ranks water first but agrees more closely with the exact Shapley order at lower ranks.

For the TabM (Table \ref{tab:ucs_tabm}), the consistency across different methods is even more significant. The rankings generated by SamplingSHAP and KernelSHAP are almost identical, differing only at ranks 13 and 14, where features 10 (\textit{scm1\_ssa}) and 12 (\textit{scm2\_ssa}) are interchanged. Meanwhile, Exact Shapley and Exact Owen again produce identical rankings. In addition, Exact CIS, ES, and FESP are completely consistent. Most methods consistently rank features 1 (\textit{age}) and 2 (\textit{sp}) as the two most important features, while features 7 (\textit{opc}), 11 (\textit{fa}), and 15 (\textit{mt\_pi}) are also frequently ranked among the most influential features.

Overall, the feature rankings generated by Sampling Owen, Exact Owen, and Exact Shapley are highly consistent, particularly for the top-ranked features, with only slight differences at the lower ranks.

\textbf{Quality of explanation results.} The values of $\mathcal{S}$ scores are summarized in Table \ref{tab:scores}. For the regression task on the UCS dataset (see Figure \ref{fig:ucs}), Sampling Owen achieves $\mathcal{S}=0.5319$ for LightGBM and $\mathcal{S}=0.4925$ for TabM. With LightGBM, it is essentially comparable to Exact Shapley and Exact Owen (both have $\mathcal{S}=0.5320$) and it demonstrates the best result   among all approximate methods, outperforming LeverageSHAP ($\mathcal{S}=0.5081$) and Original ShapG ($\mathcal{S}=0.5020$) and, by a small margin, the AFA-based methods. The Sampling Owen curve closely follows those of Exact Owen and Exact Shapley across all feature drops (see Figure \ref{fig:ucs}). With TabM, Sampling Owen ($\mathcal{S}=0.4925$) slightly exceeds Exact Owen ($\mathcal{S}=0.4912$), and it is slightly worse than the best-performing method, Original ShapG ($\mathcal{S}=0.4987$). By contrast, CIS and ES obtain the lowest scores ($\mathcal{S}=0.3521$ with LightGBM), indicating that methods based on individual feature values may ignore important relationships between features.

For the Adult Income dataset (see Figure \ref{fig:adult}), Sampling Owen achieves the highest weighted slope score among all methods for LightGBM ($\mathcal{S}=0.0732$), outperforming Exact Shapley, Exact Owen, and LeverageSHAP (all $\mathcal{S}=0.0700$). Its curve drops earlier and faster than the exact methods, indicating that it better identifies the most important features. For TabM, Improved ShapG achieves the highest score ($\mathcal{S}=0.0573$). Sampling Owen matches Exact Shapley and Exact Owen in attribution accuracy ($\mathcal{S}=0.0543$), and the curves for these three methods overlap.

For the large-scale Crime dataset (121 features, see Figure \ref{fig:crime}), exact methods are infeasible due to their high computational complexity. With LightGBM, Original ShapG achieves the highest score ($\mathcal{S}=0.2033$), and Sampling Owen is very close to the first one, being the second  ($\mathcal{S}=0.2014$). Both curves drop steeply at the beginning. With TabM, Sampling Owen achieves the best result among all XAI methods ($\mathcal{S}=0.2409$), slightly ahead of Original ShapG ($\mathcal{S}=0.2378$) and well ahead of LeverageSHAP ($\mathcal{S}=0.1991$). Its curve is the steepest overall. The rankings of top-20 features are listed in Tables \ref{tab:crime_lgbm}--\ref{tab:crime_tabm}.

To summarize, with both AI models, Sampling Owen achieves explanation quality close to or better than the exact methods on UCS and Adult Income, where the exact methods can be computed. On Crime, where they cannot, it achieves the best score with TabM and the second-best score with LightGBM.

\begin{figure}[tbp]
  \centering
  \begin{subfigure}[b]{\linewidth}
    \includegraphics[width=\linewidth]{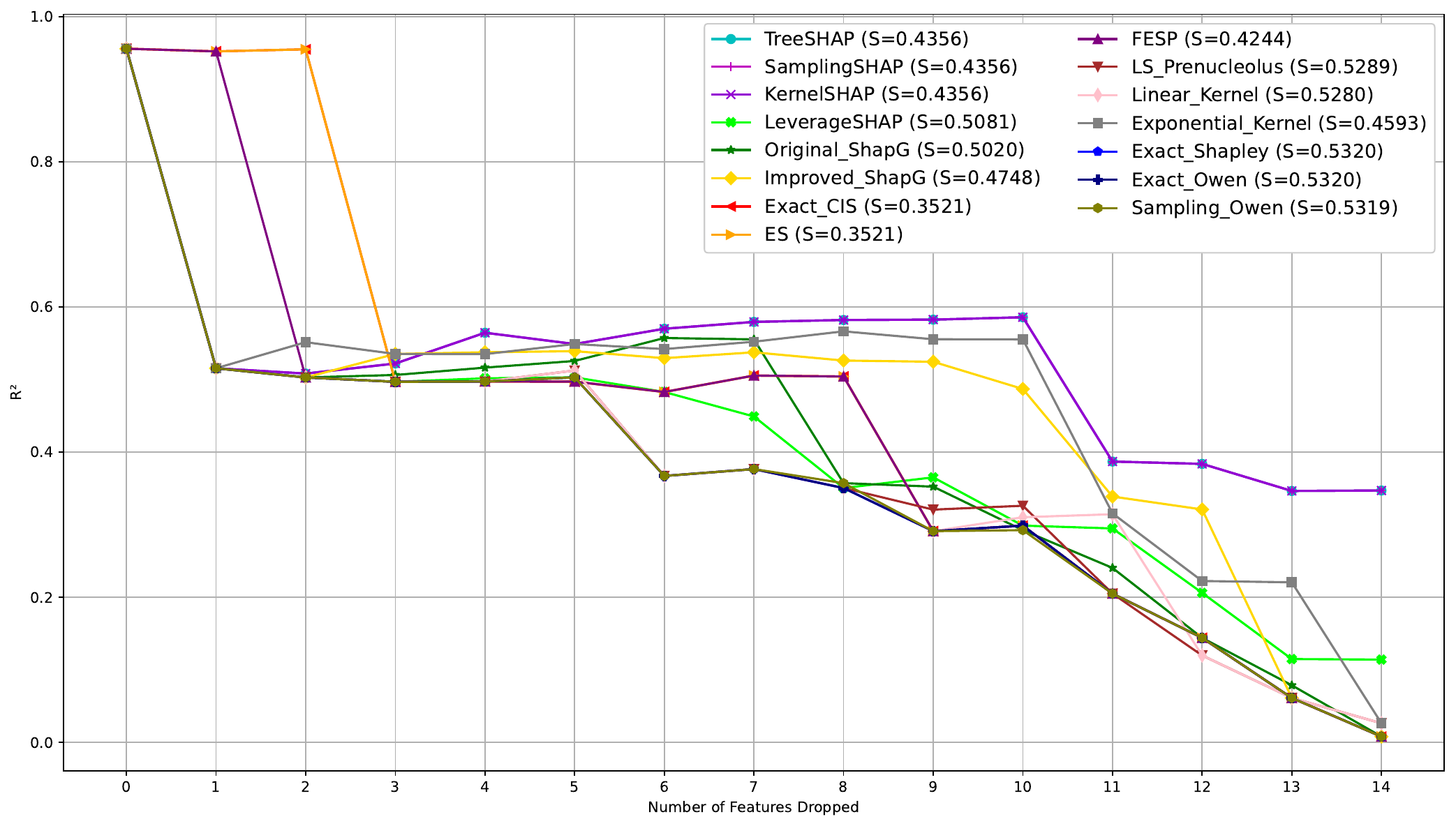}
    \caption{LightGBM}
  \end{subfigure}
  \hfill
  \begin{subfigure}[b]{\linewidth}
    \includegraphics[width=\linewidth]{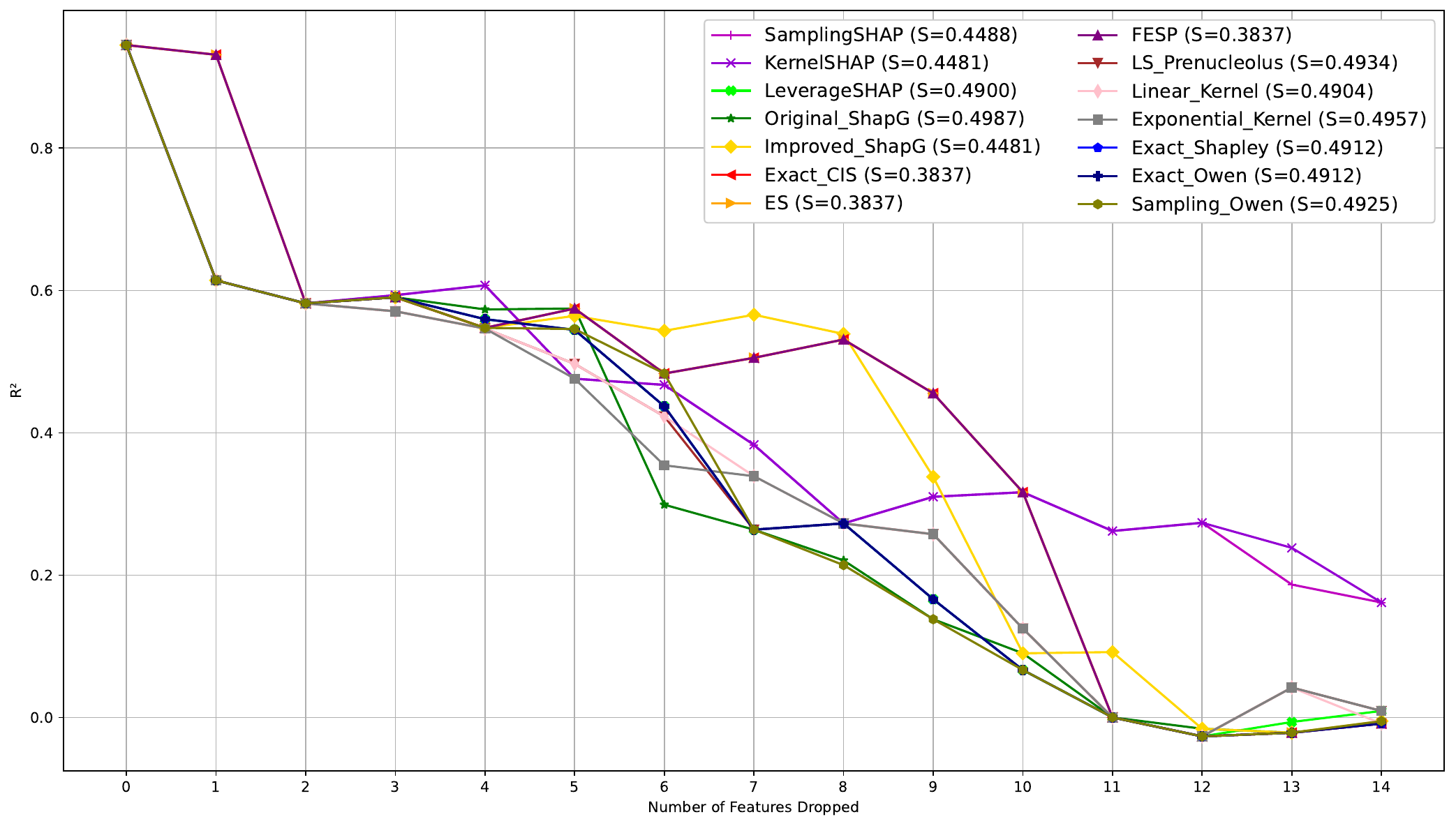}
    \caption{TabM}
  \end{subfigure}
  \caption{Feature-dropping curves for the UCS dataset (15 features, regression task). The $x$-axis represents the number of top features dropped, while the $y$-axis reports the corresponding $R^2$. Each legend entry reports the weighted slope score $\mathcal{S}$.}\label{fig:ucs}
\end{figure}

\begin{figure}[tbp]
  \centering
  \begin{subfigure}[b]{\linewidth}
    \includegraphics[width=\linewidth]{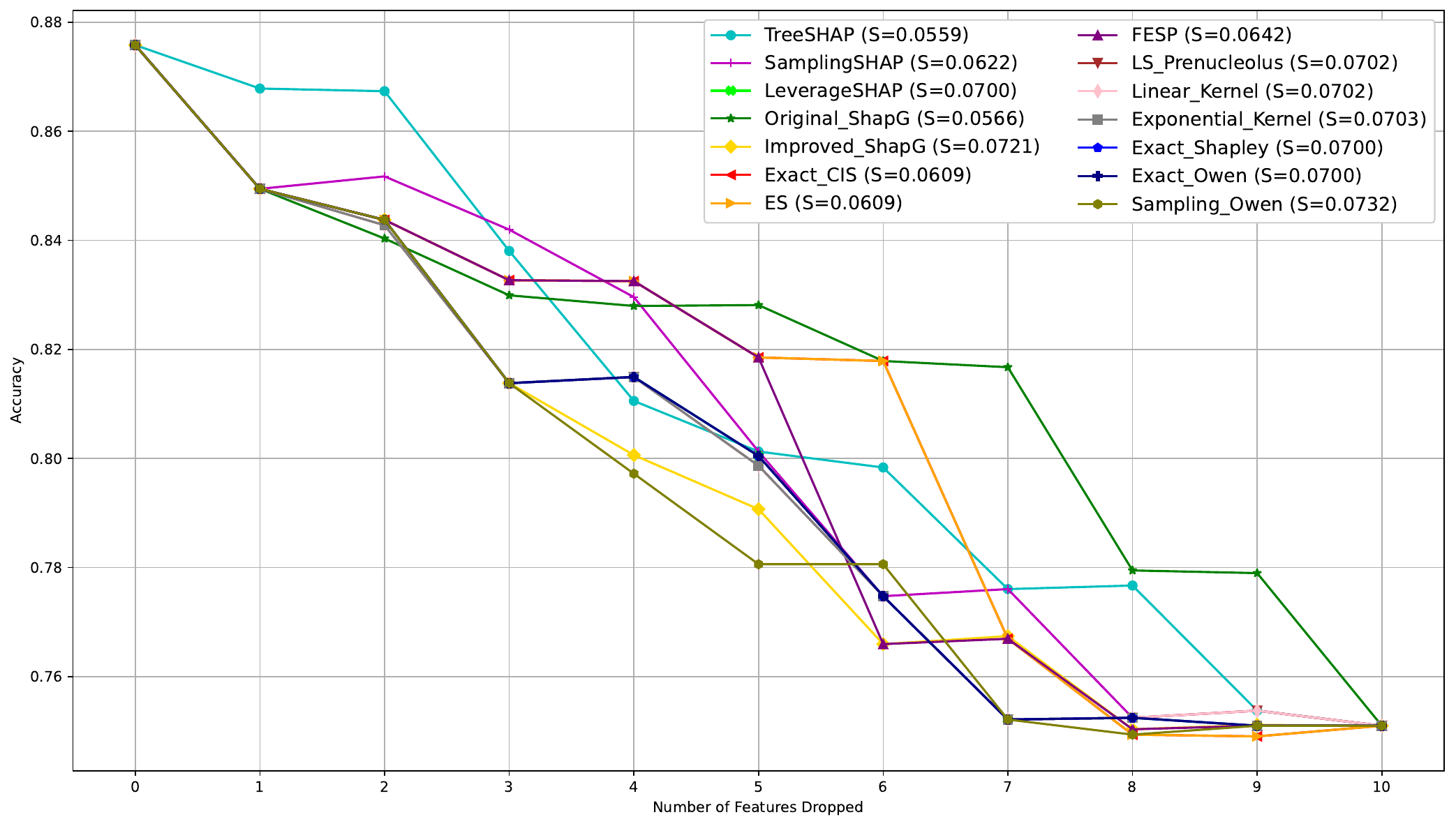}
    \caption{LightGBM}
  \end{subfigure}
  \hfill
  \begin{subfigure}[b]{\linewidth}
    \includegraphics[width=\linewidth]{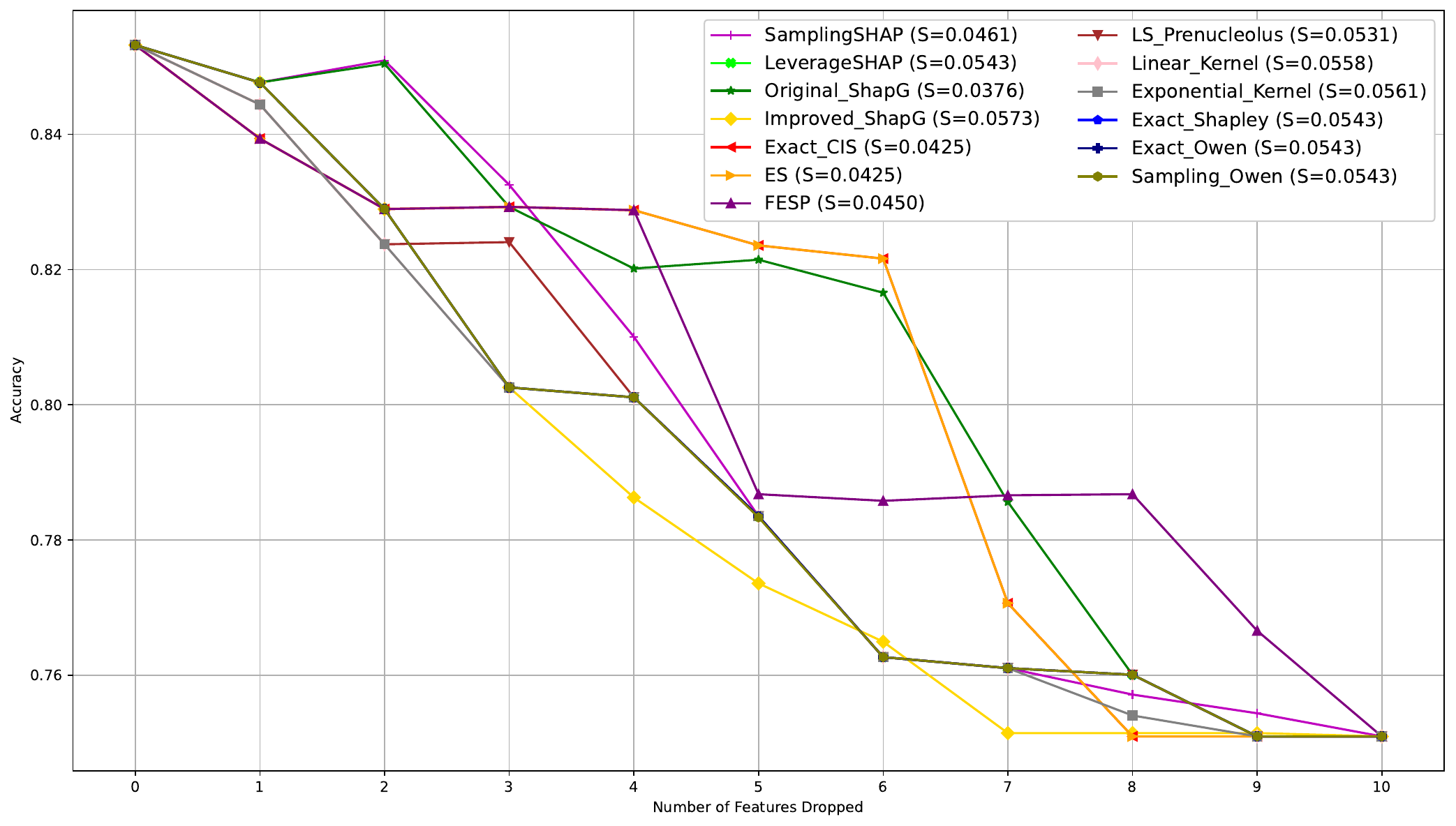}
    \caption{TabM}
  \end{subfigure}
  \caption{Feature-dropping curves for the Adult Income dataset (11 features, binary classification task).  The $x$-axis represents the number of top features dropped, while the $y$-axis reports the corresponding Accuracy. Each legend entry reports the weighted slope score $\mathcal{S}$.}\label{fig:adult}
\end{figure}

\begin{figure}[tbp]
  \centering
  \begin{subfigure}[b]{\linewidth}
    \includegraphics[width=\linewidth]{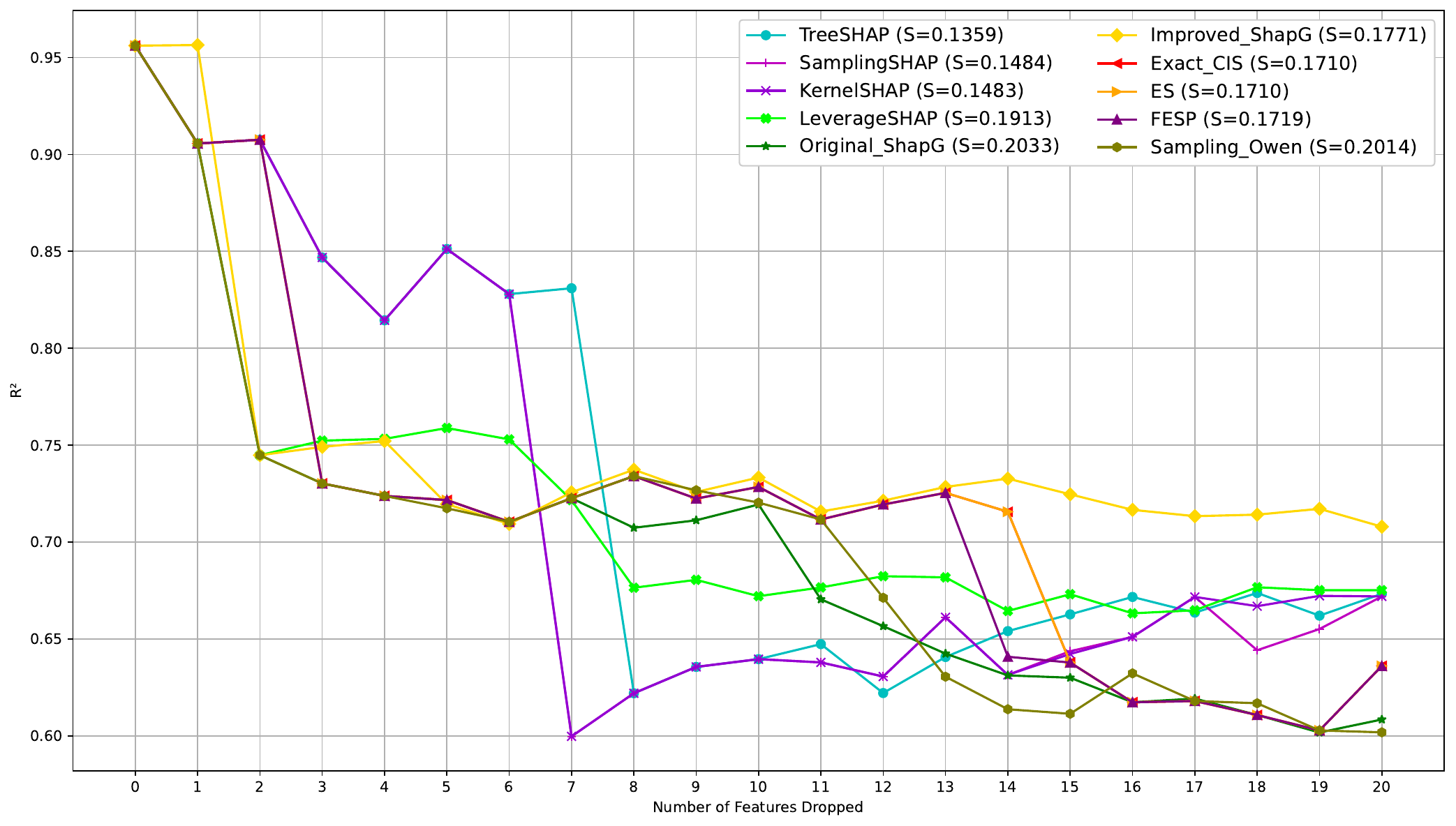}
    \caption{LightGBM}
  \end{subfigure}
  \hfill
  \begin{subfigure}[b]{\linewidth}
    \includegraphics[width=\linewidth]{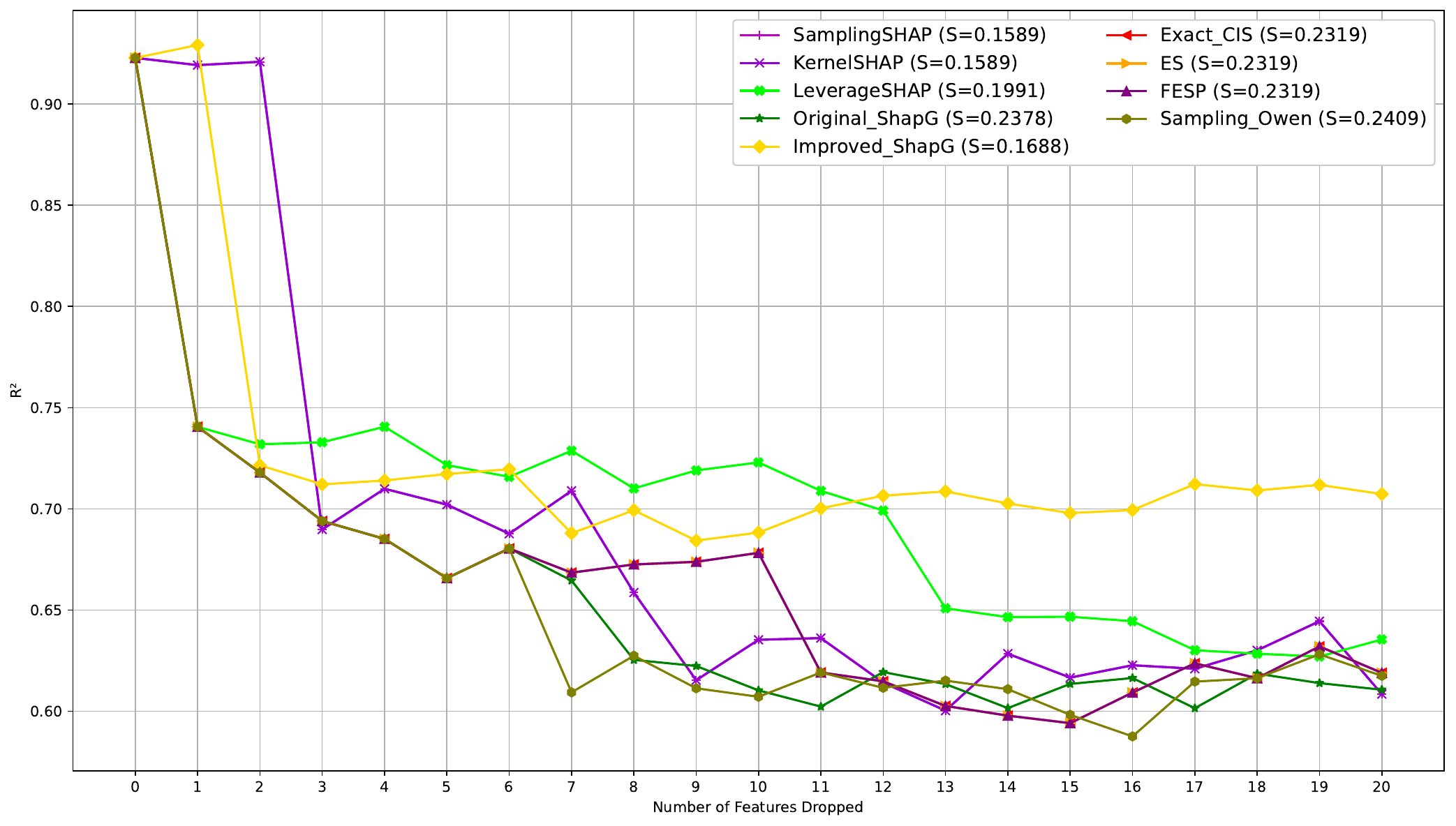}
    \caption{TabM}
  \end{subfigure}
  \caption{Feature-dropping curves for the Crime dataset (121 features, regression task).  The $x$-axis represents the number of top features dropped, while the $y$-axis reports the corresponding $R^2$. Each legend entry reports the weighted slope score $\mathcal{S}$.}\label{fig:crime}
\end{figure}

\begin{table}[!tbp]
	\centering
\caption{Weighted slope score  $\mathcal{S}$ across all XAI methods (the best score in each column is given in bold)}\label{tab:scores}%
  \small
  \setlength{\tabcolsep}{5pt}
  \begin{tabular}{l|cc|cc|cc}
    \toprule
    \diagbox{Method}{\makecell{Dataset\\Model}}
      & \makecell{UCS\\LGB}
      & \makecell{UCS\\TabM}
      & \makecell{Adult\\LGB}
      & \makecell{Adult\\TabM}
      & \makecell{Crime\\LGB}
      & \makecell{Crime\\TabM} \\
    \midrule
    TreeSHAP             & 0.4356 & ---    & 0.0559 & ---    & 0.1359 & --- \\
    SamplingSHAP         & 0.4356 & 0.4488 & 0.0622 & 0.0461    & 0.1484 & 0.1589 \\
    KernelSHAP           & 0.4356 & 0.4481 & ---    & ---    & 0.1483 & 0.1589 \\
    LeverageSHAP         & 0.5081 & 0.4900 & 0.0700 & 0.0543 & 0.1913 & 0.1991 \\
    Original ShapG      & 0.5020 & \textbf{0.4987} & 0.0566 & 0.0376 & \textbf{0.2033} & 0.2378 \\
    Improved ShapG      & 0.4748 & 0.4481 & 0.0721 & \textbf{0.0573} & 0.1771 & 0.1688 \\
    Exact CIS           & 0.3521 & 0.3837 & 0.0609 & 0.0425 & 0.1710 & 0.2319 \\
    ES                   & 0.3521 & 0.3837 & 0.0609 & 0.0425 & 0.1710 & 0.2319 \\
    FESP                 & 0.4244 & 0.3837 & 0.0642 & 0.0450 & 0.1719 & 0.2319 \\
    LS Prenucleolus     & 0.5289 & 0.4934 & 0.0702 & 0.0531 & ---    & --- \\
    Linear Kernel       & 0.5280 & 0.4904 & 0.0702 & 0.0558 & ---    & --- \\
    Exponential Kernel  & 0.4593 & 0.4957 & 0.0703 & 0.0561 & ---    & --- \\
    Exact Shapley       & \textbf{0.5320} & 0.4912 & 0.0700 & 0.0543 & ---    & --- \\
    Exact Owen          & \textbf{0.5320} & 0.4912 & 0.0700 & 0.0543 & ---    & --- \\
    Sampling Owen & 0.5319 & 0.4925 & \textbf{0.0732} & 0.0543 & 0.2014 & \textbf{0.2409} \\
    \bottomrule
  \end{tabular}
\end{table}

\begin{table}[!tbp]
	\centering
\caption{Running time across all XAI methods}\label{tab:runtime}%
  \small
  \setlength{\tabcolsep}{5pt}
  \begin{tabular}{l|rr|rr|rr}
    \toprule
    \diagbox{Method}{\makecell{Dataset\\Model}}
      & \makecell{UCS\\LGB}
      & \makecell{UCS\\TabM}
      & \makecell{Adult\\LGB}
      & \makecell{Adult\\TabM}
      & \makecell{Crime\\LGB}
      & \makecell{Crime\\TabM} \\
    \midrule
    TreeSHAP            & 0.11\,s   & ---          & 0.26\,s    & ---          & 0.21\,s    & ---           \\
    SamplingSHAP        & 11.61\,s  & 15.66\,s     & 185.89\,min & 190.97\,min & 16.77\,min & 17.45\,min    \\
    KernelSHAP          & 87.66\,s  & 185.67\,s    & $\sim$92\,h & $\sim$172\,h & 60.75\,min & 89.21\,min    \\
    LeverageSHAP        & 21.38\,s  & 360.32\,s    & 109.91\,s  & 598.64\,s    & 108.10\,s  & 583.12\,s     \\
    Original ShapG      & 7.07\,s   & 144.45\,s    & 14.68\,s   & 83.16\,s     & 77.87\,min & 14.20\,h      \\
    Improved ShapG      & 1.70\,s   & 39.83\,s     & 8.10\,s    & 43.29\,s     & 105.01\,s  & 19.47\,min    \\
    Exact CIS           & 0.98\,s   & 6.47\,s      & 0.70\,s    & 5.39\,s      & 9.01\,s    & 65.61\,s      \\
    ES                  & 0.69\,s   & 6.21\,s      & 0.65\,s    & 5.21\,s      & 8.88\,s    & 61.11\,s      \\
    FESP                & 1.77\,s   & 12.01\,s     & 1.81\,s    & 11.70\,s     & 32.90\,s   & 119.39\,s     \\
    LS Prenucleolus     & 13.54\,min & 197.74\,min & 177.65\,s  & 19.17\,min   & ---        & ---           \\
    Linear Kernel       & 13.18\,min & 198.16\,min & 177.94\,s  & 19.17\,min   & ---        & ---           \\
    Exponential Kernel  & 11.65\,min & 197.77\,min & 177.30\,s  & 19.20\,min   & ---        & ---           \\
    Exact Shapley       & 11.58\,min & 197.47\,min & 176.32\,s  & 19.12\,min   & ---        & ---           \\
    Exact Owen          & 11.51\,min & 197.61\,min & 175.67\,s  & 19.10\,min   & ---        & ---           \\
    Sampling Owen & 14.70\,s & 258.76\,s & 58.16\,s   & 289.89\,s    & 307.22\,s  & 28.48\,min    \\
    \bottomrule
  \end{tabular}
\end{table}

\textbf{Computational efficiency.} Table \ref{tab:runtime} reports the running time of all XAI methods across all datasets and AI models.

For the UCS dataset, the fastest methods are CIS and ES (under 1 s), but their $\mathcal{S}$ scores are the lowest among all XAI methods. FESP and Improved ShapG are also fast and achieve moderate scores. Exact Shapley and Exact Owen reach the highest score with LightGBM ($\mathcal{S} = 0.5320$) but require 12 min for LightGBM and over 3 hours for TabM. Sampling Owen attains almost the same $\mathcal{S} = 0.5319$ for LightGBM and achieves $\mathcal{S} = 0.4925$ for TabM. It is 47 times faster than Exact Owen for LightGBM model and 46 times faster for TabM model. LS Prenucleolus achieves almost similar scores (0.5289 and 0.4934) but takes over 13 min and 3 h. Original ShapG achieves the best score for TabM ($\mathcal{S} =0.4987$) in 144.5 s, but its score for LightGBM ($\mathcal{S} =0.5020$) is lower than for Sampling Owen.

For the Adult Income dataset, computations for the exact methods are feasible but expensive. Exact Owen and Exact Shapley take about 176 s and 19 min, with $\mathcal{S}$ scores of 0.0700 and 0.0543. CIS and ES run in below 1 s but have lower scores. Sampling Owen (58.2 s and 289.9 s) is 3 to 4 times faster than Exact Owen while achieving the highest score for LightGBM ($\mathcal{S} =0.0732$) and the same score as exact methods for TabM ($\mathcal{S} =0.0543$). Improved ShapG reaches the highest score for TabM ($\mathcal{S} =0.0573$) but a lower score for LightGBM ($\mathcal{S} =0.0721$). LeverageSHAP matches exact method scores but runs longer than Sampling Owen. SamplingSHAP requires over 3 h and KernelSHAP up to 172 h, making both impractical.

On Crime dataset with 121 features, exact methods are not applicable. Fast methods such as CIS and ES run in seconds but achieve relatively low $\mathcal{S}$ scores. Original ShapG reaches the highest score ($\mathcal{S} = 0.2033$) but takes 77.9 min for LightGBM model. Sampling Owen (5.1 min and 28.5 min) achieves $\mathcal{S} =0.2014$ for LightGBM and the best score for TabM ($\mathcal{S}=0.2409$). It is 15 times faster than Original ShapG for LightGBM with a nearly identical score, and 30 times faster for TabM while achieving a slightly higher score. 

These results show that Sampling Owen offers a strong balance between explanation quality and computational cost across considered datasets of different sizes. This balance between two criteria is well captured by Pareto frontier analysis given in the following section.

\subsection{Pareto Frontier Analysis}\label{subsec:pareto}

\begin{figure}[tbp]
  \centering
  \begin{subfigure}[b]{\linewidth}
    \includegraphics[width=\linewidth]{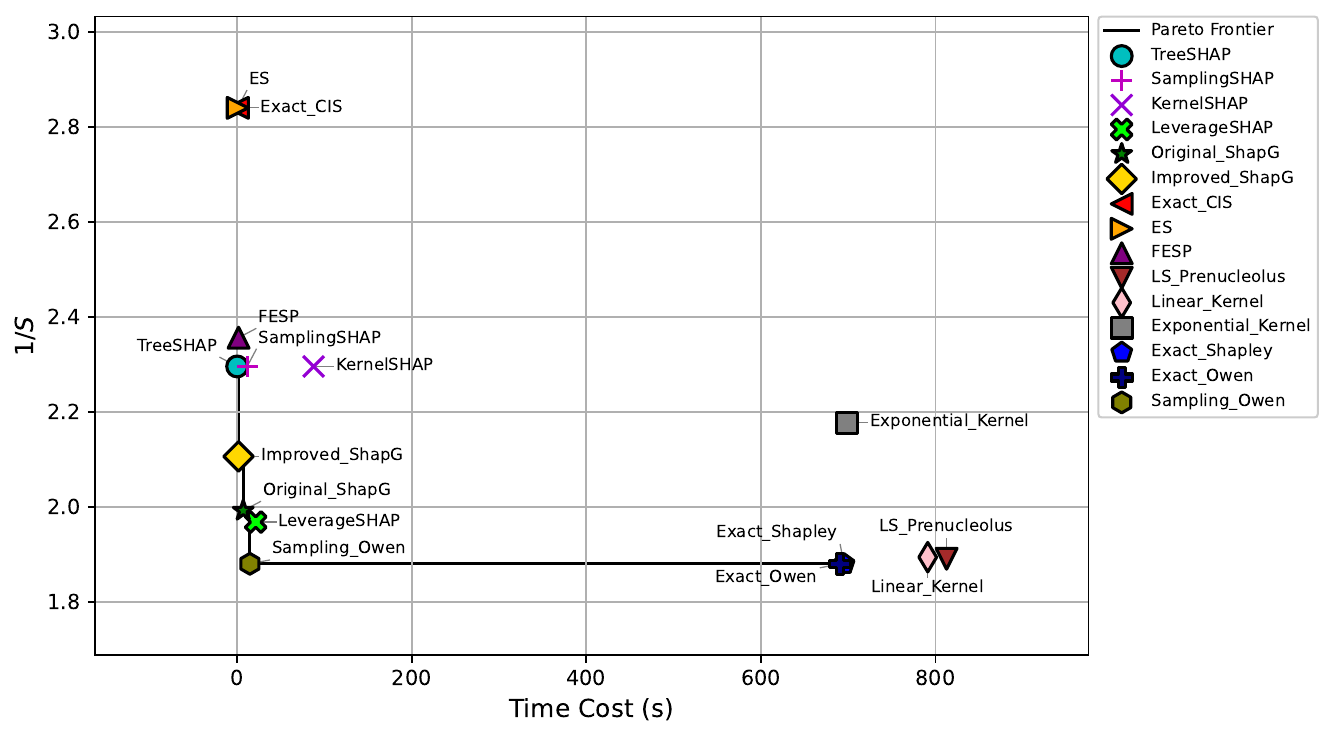}
    \caption{LightGBM}
  \end{subfigure}
  \hfill
  \begin{subfigure}[b]{\linewidth}
    \includegraphics[width=\linewidth]{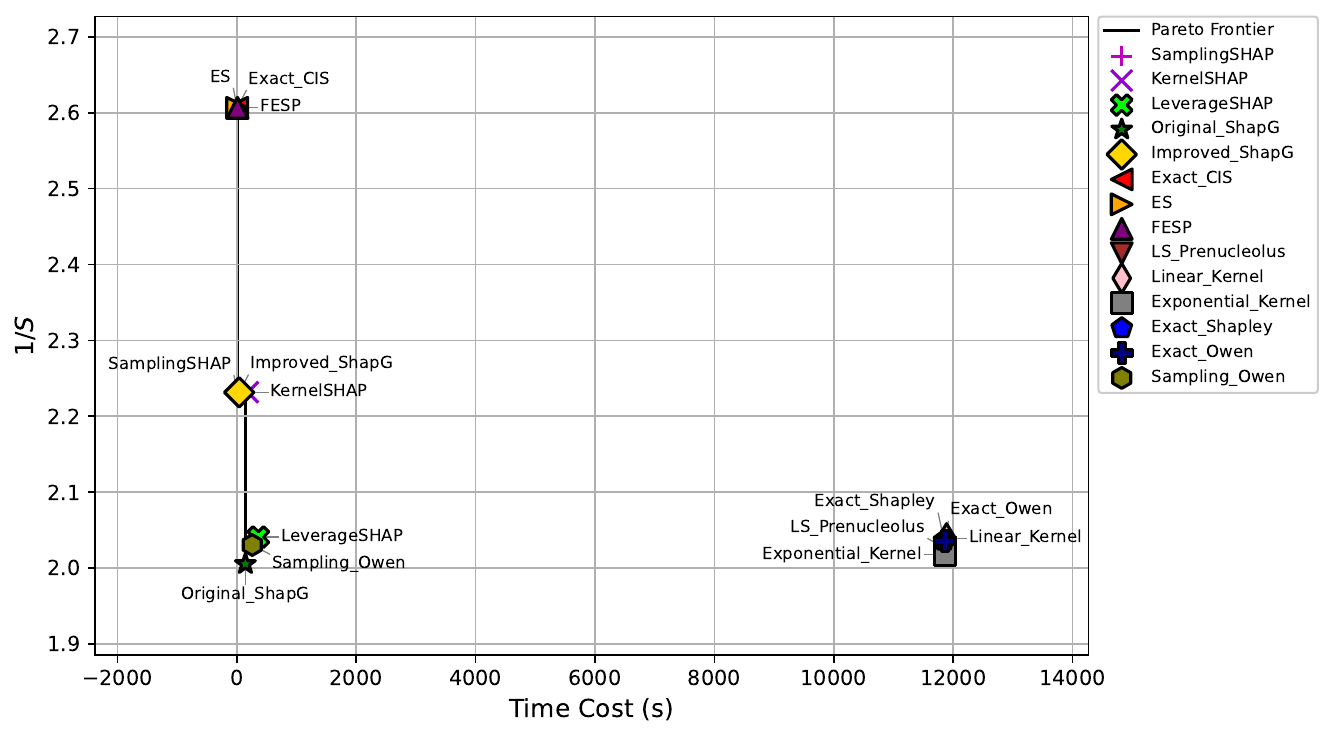}
    \caption{TabM}
  \end{subfigure}
  \caption{Pareto frontier of two-criteria evaluation (running time on $x$-axis and $1/\mathcal{S}$ on $y$-axis) for the UCS dataset (15 features, regression task). For the LightGBM model the Pareto frontier consists of TreeSHAP, Improved ShapG, Original ShapG, Sampling Owen, and Exact Owen; for the TabM model it consists of ES, SamplingSHAP and Original ShapG.}\label{fig:pareto_ucs}
\end{figure}

\begin{figure}[tbp]
  \centering
  \begin{subfigure}[b]{\linewidth}
    \includegraphics[width=\linewidth]{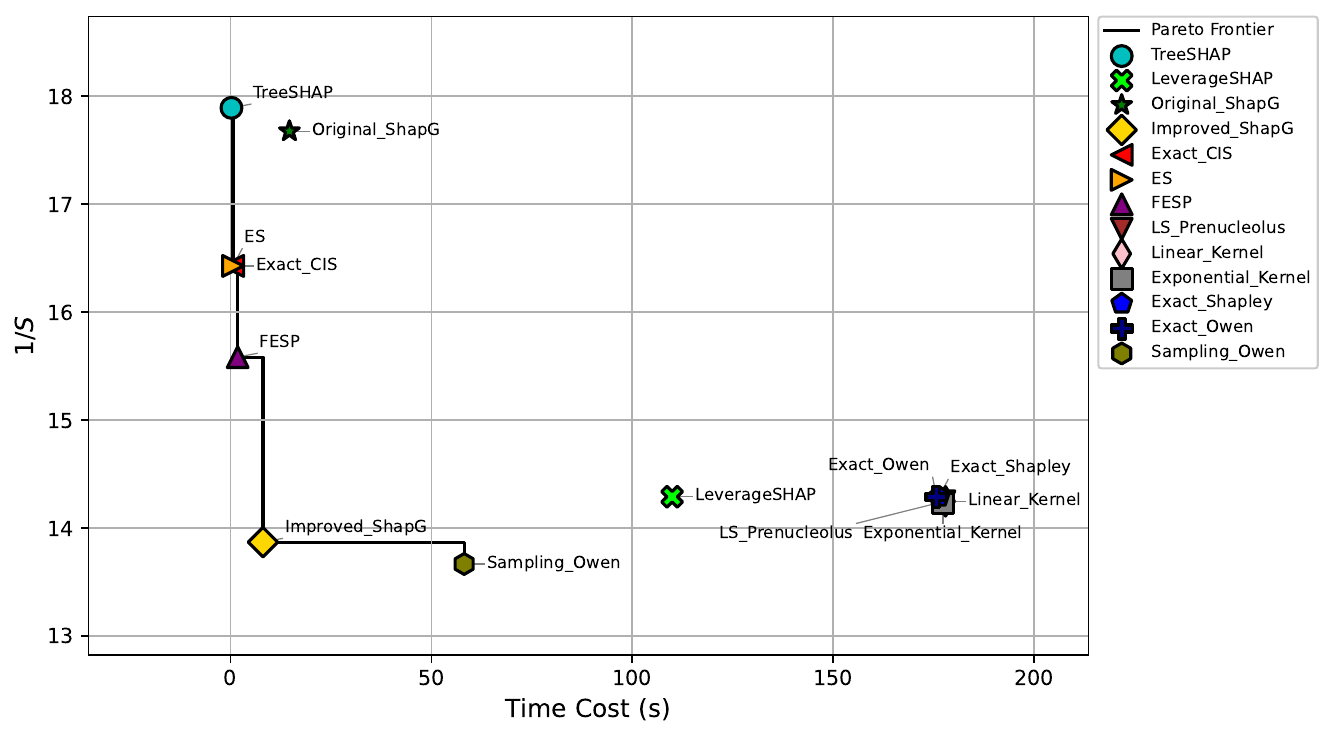}
    \caption{LightGBM}
  \end{subfigure}
  \hfill
  \begin{subfigure}[b]{\linewidth}
    \includegraphics[width=\linewidth]{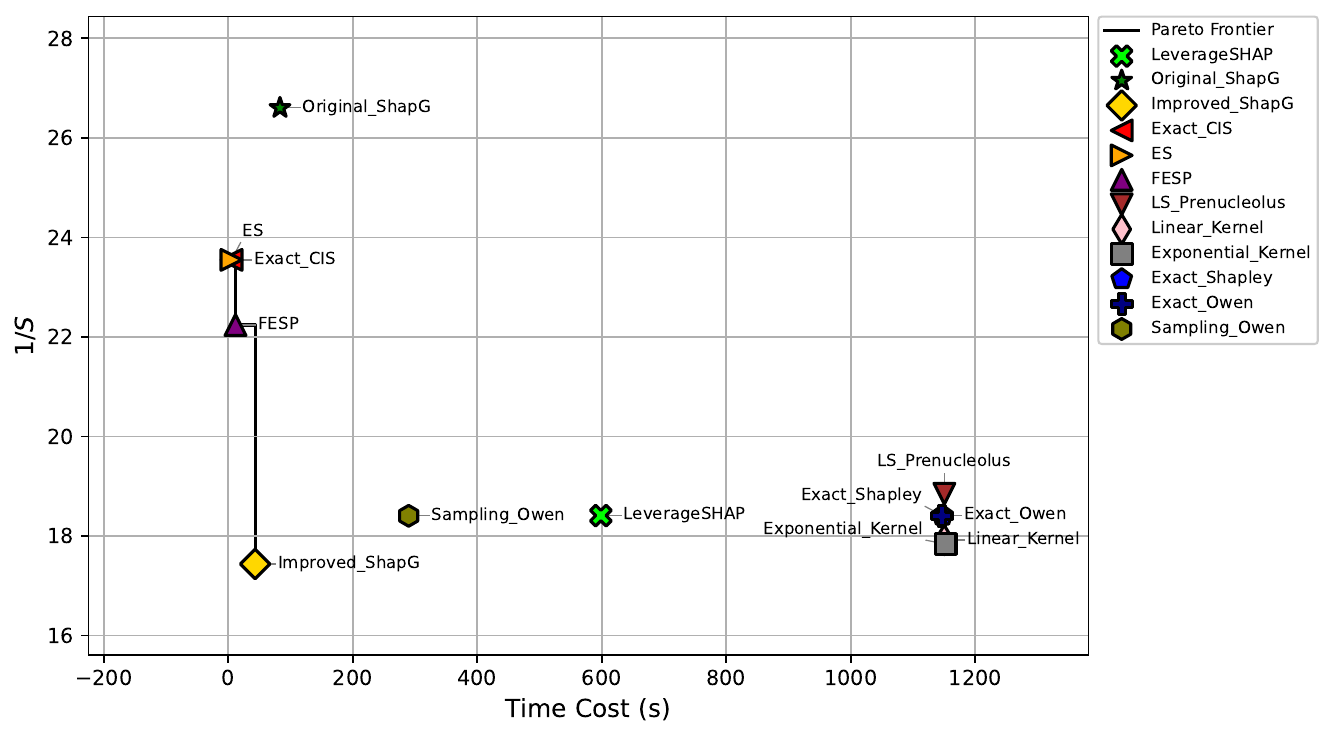}
    \caption{TabM}
  \end{subfigure}
  \caption{Pareto frontier of two-criteria evaluation (running time on $x$-axis and $1/\mathcal{S}$ on $y$-axis) for the Adult Income dataset (11 features, binary classification task). For the LightGBM model the Pareto frontier consists of TreeSHAP, ES, FESP, Improved ShapG, and Sampling Owen; for the TabM model it consists of ES, FESP and Improved ShapG.}\label{fig:pareto_adult}
\end{figure}

\begin{figure}[tbp]
  \centering
  \begin{subfigure}[b]{\linewidth}
    \includegraphics[width=\linewidth]{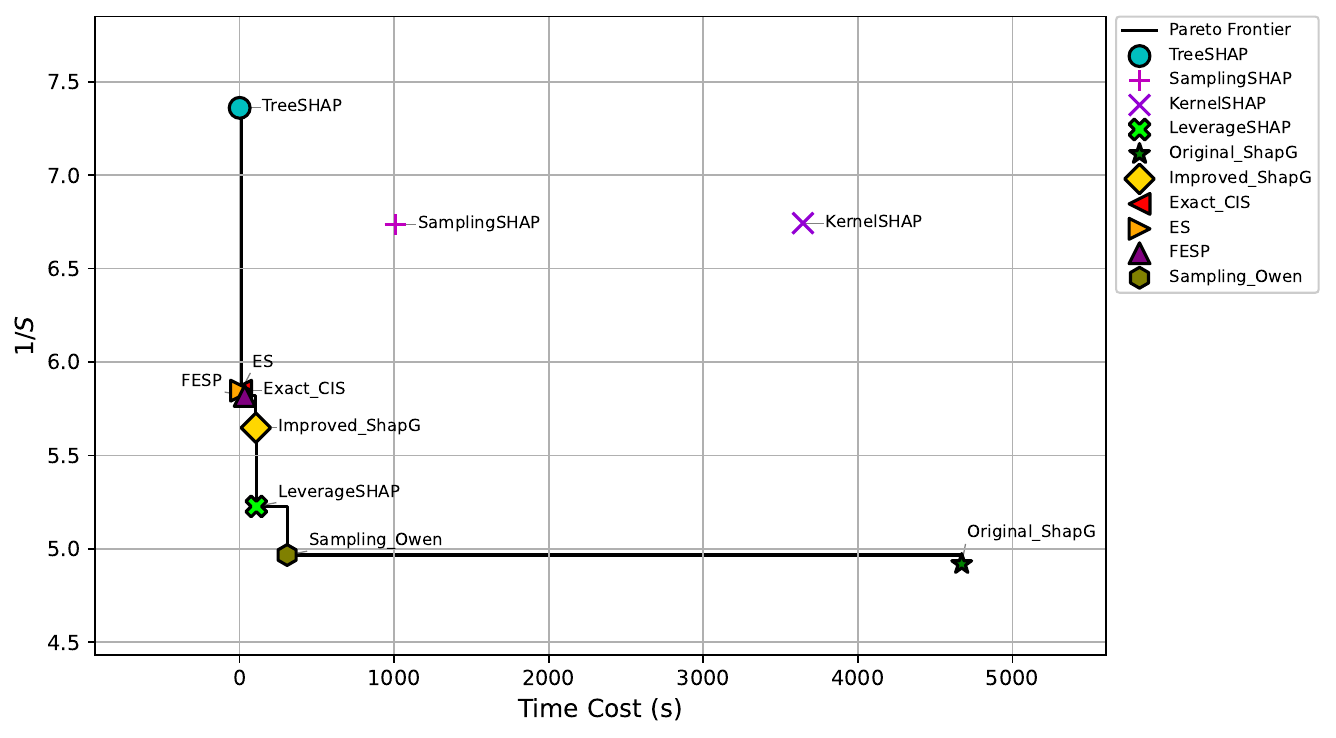}
    \caption{LightGBM}
  \end{subfigure}
  \hfill
  \begin{subfigure}[b]{\linewidth}
    \includegraphics[width=\linewidth]{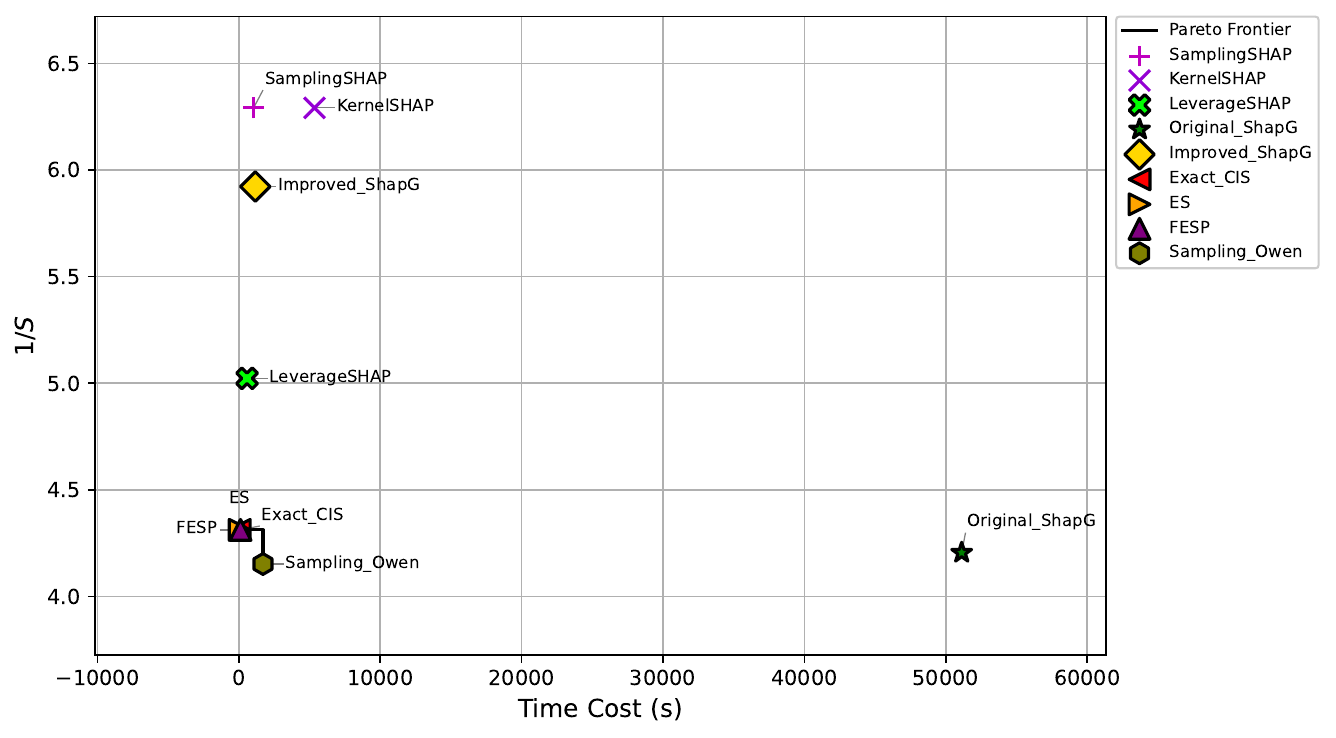}
    \caption{TabM}
  \end{subfigure}
  \caption{Pareto frontier of two-criteria method evaluation (running time on $x$-axis and $1/\mathcal{S}$ on $y$-axis) for the Crime dataset (121 features, regression task). For the LightGBM model the Pareto frontier consists of TreeSHAP, ES, FESP, Improved ShapG, LeverageSHAP, Sampling Owen and Original ShapG; for the TabM model it consists of ES and Sampling Owen.}\label{fig:pareto_crime}
\end{figure}

Pareto analysis is effective in identifying performance trade-offs in multi-objective optimization \cite{ngatchou2005pareto}. Therefore, this paper employs Pareto analysis to evaluate the performance of different XAI methods in trade-offs between explainability accuracy and computational efficiency. Figures~\ref{fig:pareto_ucs}--\ref{fig:pareto_crime} show the trade-off between computational time cost and
the inverse of the slope score, that is, $1/\mathcal{S}$ for different XAI methods. 

Since SamplingSHAP took 185.89 minutes and 190.97 minutes to run on the LGB and TabM models in the Adult Income dataset, respectively, while the running times for all other XAI methods are less than 20 minutes, the results for SamplingSHAP are not presented in the Pareto analysis to ensure the readability of the figures.

Across three datasets and two AI models, Sampling Owen is on the Pareto frontier in four out of six cases, demonstrating its ability to consistently achieve an excellent trade-off between computational efficiency and explanation quality among the evaluated XAI methods. In particular, on the Crime dataset with LightGBM, Sampling Owen achieves $\mathcal{S}$ score that is almost the same as that of the best-performing method, Original ShapG, while running about 15 times faster. This makes it a practical choice for real-world applications. Overall, the Pareto analysis further shows that Sampling Owen achieves an excellent balance between runtime and explanation quality across datasets of different sizes.

\section{Conclusions}\label{sec:conclusion}

In this paper, we propose the Leverage-Enhanced Sampling Owen Method and the corresponding algorithm for Shapley-type feature attributions via Owen's multilinear extension. The method is based on a specially composed sampling procedure which  substantially reduces computational cost compared to exact methods while maintaining explanation accuracy. We prove that the estimator is an unbiased estimator of a semivalue whose weights favor marginal contributions to the smallest and the largest coalitions, and we give these weights and the variance of the estimator in closed form.

We evaluate the method on three datasets of different sizes using LightGBM and TabM models. Sampling Owen achieves explanation quality close to or better than the exact methods on the two datasets where they can be computed, while reducing running time substantially. On the large-scale Crime dataset with 121 features where exact methods are not feasible, Sampling Owen achieves the best explanation quality for TabM and competitive results for LightGBM model, while completing in minutes compared to hours required by the best-scoring alternative.

Overall, as the Pareto analysis confirms, Sampling Owen provides a strong balance between explanation quality and computational efficiency, making it a practical choice for feature importance analysis across datasets of different scales.

We believe that extending this method to non-tabular data, such as images and text, is a promising direction for future work. Another direction is to choose the size weights of the semivalue directly for a given evaluation criterion, instead of obtaining them as a by-product of the sampling design, for example the Shapley weights when numerical agreement with the Shapley value is required.

\section*{Declarations}

\begin{itemize}
\item \textbf{Conflict of interest.} The authors have no competing interests to declare that are relevant to the content of this article.
\item \textbf{Data availability.} All datasets used in this study are publicly available. Readers can access and download them directly from the internet using the dataset names provided in the manuscript.
\item \textbf{Code availability.} The implementation of the proposed Sampling Owen method, as well as the reproduction scripts and figures reported in this paper, are publicly available at
  \url{https://github.com/LIU2216/sampling-owen}.
\end{itemize}

\bibliographystyle{elsarticle-num}

\appendix

\section{Feature Rankings}

\subsection*{Dataset: Adult Income}

Tables \ref{tab:adult_lgbm} and \ref{tab:adult_tabm} present the feature importance rankings generated by  XAI methods for the Adult Income dataset (11 features, binary classification task) for LightGBM and TabM models, respectively.

\begin{sidewaystable}[htbp]
\centering
\caption{Feature importance ranking for the Adult Income dataset (LightGBM). Entries are feature labels.$^{a}$ Merged column headers indicate that the corresponding XAI methods produce identical explanation results.}
\label{tab:adult_lgbm}
\setlength{\tabcolsep}{3pt}
\renewcommand{\arraystretch}{1.1}
\scriptsize
\begin{tabular}{l|c|c|c|c|c|c|c|c|c}
\toprule
\diagbox[width=5em,height=3em]{Rank}{Method}
 & TreeSHAP
 & \makecell{Leverage\\SHAP\\Exact Shapley\\Exact Owen}
 & \makecell{Original\\ShapG}
 & \makecell{Improved\\ShapG}
 & \makecell{Exact CIS\\ES}
 & FESP
 & \makecell{LS\\Prenucleolus\\Linear\\Kernel}
 & \makecell{Exponential\\Kernel}
 & \makecell{Sampling\\Owen}
\\
\midrule
Top 1  &  1 &  9 &  9 &  9 &  9 &  9 &  9 &  9 &  9 \\
Top 2  &  4 &  3 &  1 &  3 &  3 &  3 &  5 &  5 &  3 \\
Top 3  &  9 &  5 &  3 &  5 & 10 & 10 &  3 &  3 &  5 \\
Top 4  &  6 &  6 &  2 &  2 &  2 &  2 &  6 &  6 & 10 \\
Top 5  &  3 &  1 &  4 & 10 &  1 &  1 &  4 &  4 &  1 \\
Top 6  & 11 &  4 & 10 &  1 &  4 &  5 &  1 &  1 &  6 \\
Top 7  &  5 & 10 & 11 &  8 &  5 &  4 & 10 & 10 &  4 \\
Top 8  &  8 & 11 &  6 & 11 &  6 & 11 & 11 & 11 &  2 \\
Top 9  & 10 &  2 &  7 &  7 &  7 &  6 &  8 &  2 & 11 \\
Top 10 &  2 &  8 &  5 &  4 &  8 &  8 &  2 &  8 &  8 \\
Top 11 &  7 &  7 &  8 &  6 & 11 &  7 &  7 &  7 &  7 \\
\bottomrule
\end{tabular}
\par\medskip
{\scriptsize $^{a}$ Feature labels: 1: age; 2: workclass; 3: education\_num; 4: marital\_status; 5: occupation; 6: relationship; 7: race; 8: sex; 9: capital\_gain; 10: capital\_loss; 11: hours\_per\_week.}
\end{sidewaystable}

\begin{sidewaystable}[htbp]
\centering
\caption{Feature importance ranking for the Adult Income dataset (TabM). Entries are feature labels.$^{a}$ Merged column headers indicate that the corresponding XAI methods produce identical explanation results.}
\label{tab:adult_tabm}
\setlength{\tabcolsep}{3pt}
\renewcommand{\arraystretch}{1.1}
\scriptsize
\begin{tabular}{l|c|c|c|c|c|c|c|c|c}
\toprule
\diagbox[width=5em,height=3em]{Rank}{Method}
 & \makecell{Leverage\\SHAP\\Exact Shapley\\Exact Owen}
 & \makecell{Original\\ShapG}
 & \makecell{Improved\\ShapG}
 & \makecell{Exact CIS\\ES}
 & FESP
 & \makecell{LS\\Prenucleolus}
 & \makecell{Linear\\Kernel}
 & \makecell{Exponential\\Kernel}
 & \makecell{Sampling\\Owen}
\\
\midrule
Top 1  &  3 &  3 &  3 &  9 &  9 &  5 &  5 &  5 &  3 \\
Top 2  &  9 &  2 &  9 &  3 &  3 &  3 &  3 &  3 &  9 \\
Top 3  &  5 &  9 &  5 &  2 &  2 &  6 &  9 &  9 &  5 \\
Top 4  &  6 &  1 &  1 & 10 & 10 &  9 &  6 &  6 &  6 \\
Top 5  &  4 &  4 &  2 &  1 &  5 &  4 &  4 &  4 &  1 \\
Top 6  &  1 & 11 & 11 &  4 &  7 &  1 &  1 &  1 &  4 \\
Top 7  & 11 &  6 & 10 &  5 &  8 & 11 & 11 & 11 &  2 \\
Top 8  &  2 &  5 &  8 &  6 &  6 &  2 &  2 & 10 & 11 \\
Top 9  & 10 & 10 &  4 &  7 &  1 & 10 & 10 &  2 & 10 \\
Top 10 &  8 &  7 &  6 &  8 & 11 &  8 &  8 &  8 &  8 \\
Top 11 &  7 &  8 &  7 & 11 &  4 &  7 &  7 &  7 &  7 \\
\bottomrule
\end{tabular}
\par\medskip
{\scriptsize $^{a}$ Feature labels: 1: age; 2: workclass; 3: education\_num; 4: marital\_status; 5: occupation; 6: relationship; 7: race; 8: sex; 9: capital\_gain; 10: capital\_loss; 11: hours\_per\_week.}
\end{sidewaystable}

\subsection*{Dataset: Communities and Crime}

Tables \ref{tab:crime_lgbm} and \ref{tab:crime_tabm} represent  feature importance rankings of top-20 features generated by each XAI method on the Communities and Crime dataset (121 features, regression) for LightGBM and TabM models, respectively. Since this dataset contains 121 features, exact methods (Exact Shapley, Exact Owen, LS-Prenucleolus, Linear Kernel, Exponential Kernel) are computationally infeasible. Entries are feature labels$^{a}$.

\begin{sidewaystable}[htbp]
\centering
\caption{Importance ranking for top-20 features for the Crime dataset (LightGBM). Entries are feature labels.$^{a}$ Merged column headers indicate that the corresponding methods produce identical explanation results.}
\label{tab:crime_lgbm}
\setlength{\tabcolsep}{3pt}
\renewcommand{\arraystretch}{1.1}
\scriptsize
\begin{tabular}{l|c|c|c|c|c|c|c|c|c}
\toprule
\diagbox[width=5em,height=3em]{Rank}{Method}
 & TreeSHAP
 & \makecell{Exact\\CIS\\ES}
 & FESP
 & \makecell{Original\\ShapG}
 & \makecell{Improved\\ShapG}
 & \makecell{Leverage\\SHAP}
 & \makecell{Sampling\\SHAP}
 & \makecell{Kernel\\SHAP}
 & \makecell{Sampling\\Owen}
\\
\midrule
Top 1  &  1 &  1 &  1 &  1 &  3 &  1 &  1 &  1 &  1 \\
Top 2  &  2 &  2 &  2 &  3 &  1 &  3 &  2 &  2 &  3 \\
Top 3  & 15 &  3 &  3 &  2 &  6 & 26 & 15 & 15 &  2 \\
Top 4  & 12 &  4 &  4 &  4 &  8 & 11 & 12 & 12 &  4 \\
Top 5  & 18 &  5 &  5 &  5 &  2 & 27 & 18 & 18 &  6 \\
Top 6  & 22 &  6 &  6 &  6 &  9 & 28 & 22 & 22 &  5 \\
Top 7  & 36 &  7 &  7 &  7 & 12 &  2 &  3 &  3 &  7 \\
Top 8  &  3 &  8 &  8 & 10 & 14 & 15 & 36 & 36 &  8 \\
Top 9  & 37 &  9 &  9 &  8 & 19 & 21 & 37 & 37 & 11 \\
Top 10 &  7 & 10 & 10 & 11 & 18 & 29 &  7 &  7 &  9 \\
Top 11 & 35 & 11 & 11 & 15 &  5 & 30 &  5 &  5 & 10 \\
Top 12 &  5 & 12 & 12 & 13 & 23 &  9 & 10 & 10 & 15 \\
Top 13 & 38 & 13 & 13 & 12 & 10 & 31 & 35 & 35 & 12 \\
Top 14 & 25 & 14 & 15 & 16 & 21 & 32 & 14 & 14 & 17 \\
Top 15 & 10 & 15 & 14 &  9 & 13 & 33 & 38 &  8 & 14 \\
Top 16 & 39 & 16 & 16 & 14 &  4 &  6 &  8 & 38 & 16 \\
Top 17 & 31 & 17 & 17 & 18 & 24 & 14 & 25 & 25 & 13 \\
Top 18 &  8 & 18 & 18 & 17 & 25 & 25 & 11 & 40 & 19 \\
Top 19 & 11 & 19 & 19 & 22 & 17 & 34 & 31 & 31 & 18 \\
Top 20 & 14 & 20 & 20 & 19 & 16 & 35 & 40 & 11 & 21 \\
\bottomrule
\end{tabular}
\par\medskip
{\scriptsize $^{a}$ Feature labels: 1: assaultPerPop; 2: robbbPerPop; 3: assaults; 4: burglPerPop; 5: PctKids2Par; 6: robberies; 7: PctKidsBornNeverMar; 8: NumKidsBornNeverMar; 9: murders; 10: PctFam2Par; 11: nonViolPerPop; 12: rapes; 13: PctYoungKids2Par; 14: burglaries; 15: rapesPerPop; 16: racePctWhite; 17: PctTeen2Par; 18: murdPerPop; 19: autoTheft; 20: TotalPctDiv; 21: NumUnderPov; 22: state\_encoded; 23: PctHousNoPhone; 24: larcenies; 25: racepctblack; 26: FemalePctDiv; 27: communityName\_freq; 28: pctWSocSec; 29: pctWFarmSelf; 30: RentLowQ; 31: OwnOccHiQuart; 32: PctLargHouseFam; 33: agePct12t21; 34: PctRecImmig10; 35: larcPerPop; 36: OwnOccQrange; 37: autoTheftPerPop; 38: MedOwnCostPctIncNoMtg; 39: PctUsePubTrans; 40: whitePerCap.}
\end{sidewaystable}

\begin{sidewaystable}[htbp]
\centering
\caption{Importance ranking for top-20 features for the Crime dataset (TabM). Entries are feature labels.$^{a}$ Merged column headers indicate that the corresponding methods produce identical explanation results.}
\label{tab:crime_tabm}
\setlength{\tabcolsep}{3pt}
\renewcommand{\arraystretch}{1.1}
\scriptsize
\begin{tabular}{l|c|c|c|c|c|c}
\toprule
\diagbox[width=5em,height=3em]{Rank}{Method}
 & \makecell{Exact CIS\\ES\\FESP}
 & \makecell{Original\\ShapG}
 & \makecell{Improved\\ShapG}
 & \makecell{Leverage\\SHAP}
 & \makecell{Sampling SHAP\\Kernel SHAP}
 & \makecell{Sampling\\Owen}
\\
\midrule
Top 1  &  1 &  1 &  3 &  1 &  2 &  1 \\
Top 2  &  2 &  2 &  1 & 20 & 39 &  2 \\
Top 3  &  3 &  3 & 10 & 29 &  1 &  3 \\
Top 4  &  4 &  4 &  7 & 30 & 19 &  4 \\
Top 5  &  5 &  5 &  9 &  2 & 34 &  5 \\
Top 6  &  6 &  6 & 12 & 31 &  8 &  6 \\
Top 7  &  7 &  8 &  2 & 32 & 40 & 11 \\
Top 8  &  8 & 11 &  6 & 33 & 11 & 10 \\
Top 9  &  9 & 14 & 17 & 23 &  4 &  9 \\
Top 10 & 10 & 15 & 23 & 21 & 14 &  8 \\
Top 11 & 11 & 13 &  8 &  3 & 13 &  7 \\
Top 12 & 12 & 18 & 24 & 34 &  5 & 15 \\
Top 13 & 13 &  7 & 16 & 11 & 15 & 18 \\
Top 14 & 14 & 10 & 25 & 17 & 41 & 13 \\
Top 15 & 15 &  9 & 14 & 35 & 42 & 12 \\
Top 16 & 16 & 21 & 15 & 36 & 43 & 14 \\
Top 17 & 17 & 12 & 13 &  4 & 18 & 17 \\
Top 18 & 18 & 16 & 26 & 38 & 44 & 16 \\
Top 19 & 19 & 22 & 27 & 37 & 16 & 21 \\
Top 20 & 20 & 20 & 28 & 10 & 21 & 19 \\
\bottomrule
\end{tabular}
\par\medskip
{\scriptsize $^{a}$ Feature labels: 1: assaultPerPop; 2: robbbPerPop; 3: assaults; 4: burglPerPop; 5: PctKidsBornNeverMar; 6: PctKids2Par; 7: NumKidsBornNeverMar; 8: PctFam2Par; 9: robberies; 10: burglaries; 11: rapesPerPop; 12: rapes; 13: racePctWhite; 14: PctYoungKids2Par; 15: PctTeen2Par; 16: murdPerPop; 17: murders; 18: pctWInvInc; 19: autoTheftPerPop; 20: FemalePctDiv; 21: nonViolPerPop; 22: pctWPubAsst; 23: NumUnderPov; 24: autoTheft; 25: larcenies; 26: arsons; 27: HousVacant; 28: PctHousNoPhone; 29: pctWSocSec; 30: pctWFarmSelf; 31: OwnOccHiQuart; 32: communityName\_freq; 33: RentLowQ; 34: racepctblack; 35: PctNotHSGrad; 36: racePctHisp; 37: PctLargHouseOccup; 38: agePct12t21; 39: state\_encoded; 40: MedOwnCostPctInc; 41: larcPerPop; 42: arsonsPerPop; 43: RentHighQ; 44: pctUrban.}
\end{sidewaystable}

\clearpage
\addtocounter{page}{-1}

\end{document}